\documentclass[a4paper,11pt]{article}
\pdfoutput=1
\usepackage{jheppub}

\usepackage{tikz}

\usepackage{enumitem}
\usepackage[T1]{fontenc} 
\usepackage{floatrow}
\usepackage{appendix}
\usepackage{cancel}
\usepackage{tabularx}
\usepackage{comment}
\usepackage[normalem]{ulem}

\allowdisplaybreaks

\usepackage{epsf}
\usepackage{amsmath}
\usepackage{amsfonts}
\usepackage{amssymb}
\usepackage{psfrag,epsfig,graphicx,graphics}

\newcommand\numberthis[1][]{%
    \refstepcounter{equation}%
    \ifx#1\empty\else\label{eq:#1}\fi%
    \tag{\theequation}%
}

\usepackage{xargs} 
\usepackage[colorinlistoftodos,prependcaption,textsize=tiny]{todonotes}
\newcommandx{\MF}[2][1=]{\todo[linecolor=blue,backgroundcolor=blue!25,bordercolor=blue,#1]{#2}}

\newcommandx{\REB}[2][1=]{\todo[linecolor=black,backgroundcolor=white!25,bordercolor=black,#1]{#2}}

\newcommandx{\SWA}[2][1=]{\todo[linecolor=red,backgroundcolor=red!25,bordercolor=red,#1]{#2}}

\newcommandx{\LSZ}[2][1=]{\todo[linecolor=violet,backgroundcolor=violet!25,bordercolor=violet,#1]{#2}}

\newcommandx{\AP}[2][1=]{\todo[linecolor=green,backgroundcolor=green!25,bordercolor=green,#1]{#2}}

\providecommand{\U}[1]{\protect\rule{.1in}{.1in}}

\def\slashchar#1{\setbox0=\hbox{$#1$}
   \dimen0=\wd0
   \setbox1=\hbox{/} \dimen1=\wd1
   \ifdim\dimen0>\dimen1
      \rlap{\hbox to \dimen0{\hfil/\hfil}}
      #1
   \else
      \rlap{\hbox to \dimen1{\hfil$#1$\hfil}}
      /
   \fi}

\def\bei{\begin{itemize}}
\def\ei{\end{itemize}}

\def\beeq{\begin{eqnarray}} 
\def\beqa{\begin{eqnarray}}
\def\bea{\begin{eqnarray}}

\def\eea{\end{eqnarray}}
\def\eqa{\end{eqnarray}}
\def\eeeq{\end{eqnarray}}

\def\eqar{\end{array}}
\def\beqar{\begin{array}}

\def\beas{\begin{eqnarray*}}
\def\beqas{\begin{eqnarray*}}

\def\eqas{\end{eqnarray*}}
\def\eeas{\end{eqnarray*}}

\def\beq{\begin{equation}} 
\def\be{\begin{equation}}

\def\ee{\end{equation}}
\def\eq{\end{equation}}
\def\eeq{\end{equation}}

\def\beqd{\begin{displaymath}}
\def\eeqd{\end{displaymath}}
\def\eqd{\end{displaymath}}

\def\beeq{\begin{eqnarray}} \def\eeeq{\end{eqnarray}}

\newcommand{\fin}{\end{document}}

\title{\boldmath Deeply virtual meson production at HERA and at the EIC within the Color Glass Condensate EFT }

\author[a]{Renaud Boussarie,}
\author[b,c]{Luigi Delle Rose,}
\author[d]{Michael Fucilla,}
\author[b,c]{Alessandro Papa,}
\author[d]{Lech Szymanowski,}
\author[e]{Samuel Wallon,}

\affiliation[a]{CPHT, CNRS, Ecole Polytechnique, Institut Polytechnique de Paris, 91128 Palaiseau, France}

\affiliation[b]{Dipartimento di Fisica, Università della Calabria, Arcavacata di Rende, I-87036, Cosenza, Italy}

\affiliation[c]{INFN, Gruppo Collegato di Cosenza, Arcavacata di Rende, I-87036, Cosenza, Italy}

\affiliation[d]{National Centre for Nuclear Research (NCBJ), Pasteura 7, 02-093 Warsaw, Poland}

\affiliation[e]{Université Paris-Saclay, CNRS/IN2P3, IJCLab, 91405, Orsay, France}

\emailAdd{renaud.boussarie@polytechnique.edu}
\emailAdd{luigi.dellerose@unical.it}
\emailAdd{michael.fucilla@ncbj.gov.pl}
\emailAdd{alessandro.papa@fis.unical.it}
\emailAdd{lech.szymanowski@ncbj.gov.pl}
\emailAdd{samuel.wallon@ijclab.in2p3.fr}

\abstract{Continuing our previous study of Deeply Virtual Meson Production (DVMP) at twist-3 accuracy, we derive compact expressions for all helicity amplitudes. We perform a phenomenological analysis of the helicity-amplitude ratio $\mathcal{A}^{11}/\mathcal{A}^{00}$ and of the spin-density matrix element $r_{00}^{04}$ within the Color Glass Condensate framework. Small-$x$ evolution is incorporated by numerically solving the running-coupling-and-collinearly-improved Balitsky–Kovchegov and Balitsky–Fadin–Kuraev–Lipatov equations with the McLerran–Venugo\-pa\-lan model as the initial condition. Our exploratory phenomenological study shows that parameters fitted to inclusive DIS data capture the shape of the measured helicity observables but overestimate their magnitude, indicating the need for a dedicated fit to exclusive data. To overcome its absence in the literature, we introduce an effective phenomenological matching prescription. Our results are then compared to HERA data, and predictions are presented for electron–lead collisions at the future Electron–Ion Collider. We discuss the impact of non-linear effects at low photon virtuality and the role of genuine higher-twist contributions associated with light vector meson distribution amplitudes, corresponding to higher-Fock-state components of the projectile.}

\begin{document} 
\maketitle
\flushbottom

\section{Introduction}
\label{sec:intro}

The exclusive production of light vector mesons in deep inelastic scattering (DIS) has played a central role in revealing the dynamics of QCD at high energies. 
The HERA measurements of $e^{\pm}p$ collisions~\cite{H1:1995cha,H1:1997bdi,H1:2006uea,H1:2006zyl,Aaron:2010aa,H1:2012pbl,ZEUS:1995sar,ZEUS:1997fox,ZEUS:1998rvb,ZEUS:2005vwg,ZEUS:2008xhs,ZEUS:2004luu,H1:2012xlc}
established that diffractive events constitute a significant fraction of the total DIS cross section, and they provided early indications of gluon saturation at small Bjorken-$x$~\cite{GolecBiernat:1998js,Golec-Biernat:1999qor}.  
In this kinematic domain, the proton behaves as a dense gluonic system whose evolution can be described within the Color Glass Condensate (CGC) effective theory~\cite{Mueller:1993rr,Mueller:1994jq,Mueller:1994gb,Chen:1995pa,Kovchegov:1999yj,Kovchegov:1999ua,Balitsky:1995ub,Balitsky:1998kc,Balitsky:1998ya,Balitsky:2001re,JalilianMarian:1997jx,JalilianMarian:1997gr,JalilianMarian:1997dw,JalilianMarian:1998cb,Kovner:2000pt,Weigert:2000gi,Iancu:2000hn,Iancu:2001ad,Ferreiro:2001qy}, 
a framework that captures the onset of gluon saturation and describes the small-$x$ part of the hadronic wave function as a highly occupied, strongly fluctuating ensemble of color fields.  
In this picture, the target at high energy resembles a disordered color medium, where multiple scattering and non-linear gluon recombination processes become essential, and coherent interactions with these intense gauge fields dominate the dynamics. The experimental identification of this regime is one of the most intriguing prospects for modern high-energy colliders, and constitutes a central pillar of the physics program of the future Electron--Ion Collider~\cite{Boer:2011fh}. \\

However, reaching quantitative accuracy within the CGC framework requires the systematic inclusion of several subdominant effects.  
A first and essential ingredient is the incorporation of next-to-leading order corrections, which improve the stability of the perturbative expansion and reduce the scheme and scale dependence of theoretical predictions.  
In particular, running-coupling contributions~\cite{Kovchegov:2006vj}, full NLO corrections to the small-$x$ evolution kernel~\cite{Balitsky:2008zza,Balitsky:2013fea,Grabovsky:2013mba,Balitsky:2014mca,Kovner:2013ona,Lublinsky:2016meo,Caron-Huot:2015bja} and NLO corrections to the impact factors~\cite{Balitsky:2012bs,Beuf:2022ndu,Beuf:2022kyp,Beuf:2024msh,Roy:2019hwr,Caucal:2021ent,Caucal:2022ulg,Boussarie:2014lxa,Boussarie:2016ogo,Taels:2022tza,Bergabo:2022zhe,Bergabo:2022tcu,Iancu:2022gpw,Caucal:2024nsb,Fucilla:2023mkl,Fucilla:2022wcg,Boussarie:2016bkq,Mantysaari:2022bsp,Mantysaari:2022kdm}, constitute indispensable steps toward a controlled phenomenology of dense gluonic matter. A second class of refinements involves sub-eikonal corrections~\cite{Altinoluk:2014oxa,Altinoluk:2015gia,Altinoluk:2015xuy,Agostini:2019avp,Agostini:2019hkj,Altinoluk:2020oyd,Altinoluk:2021lvu,Altinoluk:2022jkk,Altinoluk:2023qfr,Agostini:2024xqs,Chirilli:2018kkw,Chirilli:2021lif,Kovchegov:2015pbl,Kovchegov:2019rrz,Cougoulic:2022gbk,Altinoluk:2025ang,Altinoluk:2025ivn}, which capture $1/W^2$-suppressed interactions, where $W^2$ stands for center-of-mass energy, absent in the strict eikonal limit. Finally, as highlighted in~\cite{Boussarie:2024pax}, since the saturation scale is not parametrically large in present and near-future experimental conditions, power-suppressed contributions proportional to inverse powers of the hard scale—{\it i.e.}\ higher-twist effects—may also induce sizeable modifications to exclusive and diffractive observables.  
This makes the systematic control of all these subleading contributions a prerequisite for precision phenomenology at small~$x$.

\vspace{0.2cm}

Beyond providing precision corrections, the inclusion of higher-twist effects can significantly enlarge the set of observables that can be meaningfully accessed. An emblematic example is represented by the process which is the object of this study, the exclusive electroproduction of light vector mesons, a.k.a. Deeply Virtual Meson Production (DVMP), which involves the subprocess
\begin{equation}
    \gamma^{(*)} (q) \; P (p_t) \to M(p_M )\, P(p_t') \; ,
\end{equation}  
where $M= \{ \rho,\phi,\omega \}$. This process has been extensively investigated both in forward electroproduction~\cite{Chekanov:2007zr,Aaron:2009xp} 
and in photoproduction at large momentum transfer~\cite{Breitweg:1999jy,Chekanov:2002rm,Aktas:2006qs,Aaron:2009xp} at HERA, 
and is known to provide access to the transverse spatial structure of the proton and to the dynamics of color-singlet exchange~\cite{Ivanov:1998gk,Ivanov:2000uq,Munier:2001nr,Forshaw:2001pf,Enberg:2003jw,Poludniowski:2003yk}. At HERA experiments~\cite{Aaron:2009xp}, the complete spin-density matrix of the DVMP has been measured, providing detailed information of all helicity amplitudes. Unfortunately, from a theoretical perspective, a complete description of the spin-density matrix is a highly challenging task.  
At moderate energies, one could attempt to describe the process within the framework of collinear factorization~\cite{Brodsky:1994kf,Frankfurt:1995jw,Collins:1996fb,Radyushkin:1997ki}.  
However, collinear factorization is rigorously justified only for the twist-2 longitudinal channel, while transitions involving transversely polarized mesons at twist~3 are plagued by end point singularities~\cite{Mankiewicz:1999tt,Anikin:2002wg}. \\

Various approaches, based on different theoretical ideas, have been developed to address the challenges of describing higher-twist contributions. One such approach relies on improved collinear approximations~\cite{Li:1992nu,Vanderhaeghen:1999xj,Goloskokov:2005sd,Goloskokov:2006hr,Goloskokov:2007nt}. However, in the high-energy kinematics relevant for HERA and the future EIC, a more natural description emerges from high-energy formulations~\cite{Ivanov:1998gk,Anikin:2009hk,Anikin:2009bf,Anikin:2011sa,Besse:2012ia,Besse:2013muy,Bolognino:2018rhb,Bolognino:2019pba,Bolognino:2021niq}, where the end-point singularities are naturally regularized by non-zero transverse momenta of $t$-channel gluons. In this case, the gluon distribution is described in terms of momentum-space unintegrated gluon density~\cite{Anikin:2011sa,Ivanov:2000cm,Hentschinski:2012kr,Golec-Biernat:1998zce,Watt:2003mx,Boroun:2024cmv,Boroun:2025pea}. Especially for the EIC physics case, a very exciting perspective is the description of the exclusive diffractive meson production at high-energy, including the saturation regime. A first synthesis of these ideas at small-$x$ was achieved in Ref.~\cite{Boussarie:2016bkq}, 
where exclusive longitudinal meson production was computed at NLO within a saturation framework in terms of leading-twist DAs.  
In subsequent works~\cite{Boussarie:2024pax,Boussarie:2024bdo}, the extension of this program to twist-3 observables was started, providing, at the LO, a complete description of the $\gamma^{(*)} \rightarrow M(\rho, \phi, \omega)$ impact factor, for arbitrary photon/meson polarization and in fully general kinematics, including both the genuine and the kinematic higher-twist effects.  \\

In this paper, we initiate a phenomenological exploration of the spin-density matrix for exclusive $\rho$-meson leptoproduction within the novel framework proposed in~\cite{Boussarie:2024pax,Boussarie:2024bdo}. In particular, we investigate to what extent dipole amplitudes constrained by inclusive DIS can describe exclusive helicity observables and assess their sensitivity to the initial condition and small-$x$ evolution prescription. The paper is organized as follows. Section 2 introduces the general formalism and describes all ingredients entering the numerical analysis, {\it i.e.} target matrix elements and collinear distribution amplitudes parametrizations. Appendix \ref{App:RenGroup} is devoted to the renormalization group evolution of the latter. Section 3 presents the two- and three-body helicity amplitudes. These amplitudes—derived from the original expressions in Refs.~\cite{Boussarie:2024pax,Boussarie:2024bdo} by employing a dipole approximation and an expansion around the forward direction—are recast into a form better suited for numerical implementation. Their derivation is lengthy and algebraically intricate, so the appendix~\ref{App:DerivAmp} is entirely devoted to this calculation. Section 4 contains the numerical results, and Section 5 summarizes our findings. 

\section{Theoretical set-up}

\subsection{CGC effective framework}

Throughout this work, we employ the semi-classical effective description of QCD at small~$x$ \cite{McLerran:1993ka,McLerran:1993ni,McLerran:1994vd,Balitsky:1995ub,Balitsky:1998kc,Balitsky:1998ya,Balitsky:2001re}. For this purpose, we introduce a light-cone basis defined by two null vectors $n_1$ and $n_2$, which specify the $+/-$ directions and satisfy $n_1 \cdot n_2 = 1$. 
The Sudakov decomposition of an arbitrary four-vector $k$ reads\footnote{
Transverse momenta in Euclidean space are denoted in boldface, while Minkowskian transverse components carry a $\perp$ subscript.
}
\begin{equation}
  k^\mu = k^+ n_1^\mu + k^- n_2^\mu + k_\perp^\mu \, .  
\end{equation}
  
In this approach, the gluon field $\mathcal{A}(k)$ is separated into a classical background field $b(k)$, dominated by small-$k^+$ modes, and quantum fluctuations $A(k)$ associated with large-$k^+$ modes.  
After performing a large boost from the target rest frame to the projectile frame, the background field takes the form
\begin{equation}
    b^\mu(x) = b^-(x^+, x_\perp)\, n_2^\mu 
    = \delta(x^+)\, \mathbf{B}(\boldsymbol{x})\, n_2^\mu \, ,
\end{equation}
corresponding to the standard \emph{shockwave approximation}. 
In this configuration, the fast projectile propagates predominantly along the $n_1$ (``$+$'') direction, while the target moves along $n_2$ (``$-$''). 
It is therefore convenient to work in the light-cone gauge
\[
A \cdot n_2 = 0 \, ,
\]
which provides the natural setting for the CGC formalism. 
Resumming the multiple scatterings of a fast projectile parton off the background field leads to a light-like Wilson line localized at $z^- = 0$,
\begin{equation}
   W_{\boldsymbol{z}} 
   = \mathcal{P} \exp\!\left( i g \int dz^+\, b^{-a}(z)\, \mathbf{T}^a_r \right) ,
   \label{Eq:GenericWilsonLine}
\end{equation}
where $\mathcal{P}$ denotes path ordering along the $+$ direction and $\mathbf{T}^a_r$ is the color generator in the representation $r$.  
In the following, $V_{\boldsymbol{z}}$ and $U_{\boldsymbol{z}}$ denote Wilson lines in the fundamental ($\mathbf{T}^a_r \to t^a_{ij}$) and adjoint ($\mathbf{T}^a_r \to -i f^{abc}$) representations, respectively. Within the CGC formalism, the scattering amplitude is expressed as the convolution of a perturbatively calculable projectile impact factor with a non-perturbative target matrix element built from Wilson-line operators evaluated between proton (or nuclear) states.

\subsection{Kinematics, frame and polarization vectors}

We denote by $q$ the four-momentum of the incoming photon and by $p_t$ the four-momentum of the target proton. The photon momentum admits the Sudakov decomposition
\begin{gather}
    q^{\mu} = q^{+} n_1^{\mu}
    - \frac{Q^{2}-\boldsymbol{q}^2}{2 q^{+}}\, n_2^{\mu}
    + q_{\perp}^{\mu} \; .
\end{gather}
The produced meson momentum can be written as
\begin{gather}
    p_M^{\mu} = p_M^+ n_1^{\mu}
    + \frac{\boldsymbol{p}_M^2 + m_M^2}{2 p_M^+} n_{2}^{\mu}
    + p_{M \perp}^{\mu} \; .
\end{gather}
Conversely, the target proton moves ultra-relativistically in the opposite direction with a dominant $-$ component,
\begin{gather}
      p_{t}^{\mu}
    = p_{t}^{-}\, n_2^{\mu} \; ,
\end{gather}
where its mass has been neglected and we have chosen $p_{t\perp}=0_\perp$. In the high-energy limit we use
\begin{gather}
    s = (q+p_t)^2 = (p_M+p_t')^2 
    \simeq 2 q \cdot p_t 
    = 2 p_M \cdot p_t' \; .
\end{gather}
After the interaction, the outgoing proton acquires a transverse momentum,
\begin{equation}
    p_t'^{\mu}
    = p_{t}^{-}\, n_2^{\mu}
    + \Delta_{\perp}^{\mu}
    = p_{t}^{-}\, n_2^{\mu}
    + q_{\perp}^{\mu}
    - p_{M\perp}^{\mu} \, .
\end{equation}
The amplitudes derived in \cite{Boussarie:2024bdo} were obtained in the general frame described above. For the purposes of the present analysis, however, it is convenient to adopt a less general frame\footnote{Throughout the paper, we keep the derivation fully general; however, the final expressions for the amplitudes used in the numerical implementation are presented in this specific frame.}, in which the incoming photon carries no transverse momentum in the Sudakov sense:
\begin{equation}
    q_{\perp} = 0_{\perp} \; , 
    \qquad 
    \Delta_{\perp} = -p_{M \perp} \; .
\end{equation}

In this frame, the polarization vectors are chosen according to the experimental convention of~\cite{HERMES:2010xol}, also adopted in Ref.~\cite{Anikin:2011sa}. 
The photon polarization vectors read
\begin{equation}
    \varepsilon_{\gamma L}^{\mu}
    = \frac{1}{Q}
      \left(
      q^{\mu}
      + \frac{2Q^2}{s} p_t^{\mu}
      \right),
    \qquad
    \varepsilon_{\gamma T}^{(\pm)}
    = \varepsilon_{\gamma \perp}^{(\pm)}
    = \frac{1}{\sqrt{2}} (0, \mp 1, -i, 0)
    \equiv (0, \boldsymbol{\varepsilon}_{\gamma}^{(\pm)}, 0) \, .
\end{equation}
The meson polarization vectors are defined as
\begin{equation}
    \varepsilon_{M L}^{\mu}
    = \frac{1}{m_M}
      \left(
      p_M^{\mu}
      - \frac{2 m_M^2}{s} p_{t'}^{\mu}
      \right),
    \qquad
    \varepsilon_{M T}^{(\pm)\mu}
    = \varepsilon_{M \perp}^{(\pm)\mu}
      - \frac{\varepsilon_{M \perp}^{(\pm)} \cdot p_{M \perp}}{p_M^+}
        n_2^{\mu} ,
\end{equation}
with
\begin{equation}
    \varepsilon_{M \perp}^{(\pm)}
    = \frac{1}{\sqrt{2}} (0, \mp 1, -i, 0)
    \equiv (0, \boldsymbol{\varepsilon}_M^{(\pm)}, 0) \; .
\end{equation}
Since we are not in strictly forward kinematics with $\Delta_{\perp}=0$, the transverse polarization vector of the meson is not purely transverse in the Sudakov sense. 
However, the induced non-transverse component does not contribute to the scattering amplitude. 
The transverse polarization vectors of the photon and meson satisfy the orthonormality relations
\begin{equation}
   \boldsymbol{\varepsilon}_{\gamma}^{(+)} \cdot \boldsymbol{\varepsilon}_M^{(+) *}
   =
   \boldsymbol{\varepsilon}_{\gamma}^{(-)} \cdot \boldsymbol{\varepsilon}_M^{(-) *}
   = 1 ,
   \qquad
   \boldsymbol{\varepsilon}_{\gamma}^{(-)} \cdot \boldsymbol{\varepsilon}_M^{(+) *}
   =
   \boldsymbol{\varepsilon}_{\gamma}^{(+)} \cdot \boldsymbol{\varepsilon}_M^{(-) *}
   = 0 \; .
\end{equation}

\subsection{BK and BFKL evolution of the dipole operator}
In this section, we describe the parametrization of the target matrix element and its small-$x$ evolution under the non-linear BK equation. Our definition of the forward dipole matrix element, at rapidity separation $\eta$, is
\begin{gather}
\mathcal{U} (\mathbf{r}, \eta)  = \lim_{p_t' \rightarrow p_t} \frac{ \langle P(p_{t'})| \mathcal{U}_{\boldsymbol{r}, \boldsymbol{0}}^{\eta} |P(p_t) \rangle  }{ 2 p_t^{-} (2 \pi) \delta(p_{t'}^{-}-p_{t}^{-})  (2 \pi)^2 \delta^2 (\mathbf{p}_{t'}-\mathbf{p}_{t} )}  \; ,
\label{Eq:ForwDipoleMatrixEle}
\end{gather}
where the dipole operator is defined as
\begin{gather}
  \mathcal{U}_{\boldsymbol{z}_1, \boldsymbol{z}_2}^{\eta} =  1 - \frac{1}{N_c} {\rm tr} \left( V_{\boldsymbol{z}_1}^{\eta} V_{\boldsymbol{z}_2}^{\eta \dagger} 
 \right) \; .
\end{gather}

In the numerical analysis, we allow for an ambiguity in the rapidity scale and define
\begin{equation}
    \eta=\ln \left(\frac{k}{\,x_{\rm Bj}} \right) \; ,
\end{equation}
while the initial condition is imposed at
$\eta_0=\ln(1/x_0)$. We consider fits corresponding to two different choices of the initial evolution point, $x_0=0.01$ and $x_0=1$. The corresponding evolution interval is therefore
\begin{equation}
    \eta-\eta_0=\ln \left( \frac{k \; x_0}{ x_{\rm Bj}} \right) .
\end{equation}
The parameter $k$ is introduced as an effective phenomenological parameter. As we will show, both the running-coupling and running-coupling-and-collinear-improved BK, with the parameters from the available fits, do not provide an optimal description of the data for $k=1$. We find that the choice $k=4$, in the absence of collinear improvement, provides a very good description of the exclusive HERA data. This value should be regarded as a phenomenological rapidity-scale matching. As discussed below, the present analysis indicates that a dedicated refit of the dipole initial condition to exclusive observables is ultimately required. The same rapidity is used for the 2-body and 3-body Fock-state contributions.
Indeed, within the collinear DVMP framework adopted here, the partons in both
Fock components form a collinear system that hadronizes into the same meson
and do not introduce distinct final-state invariant-mass scales. In the
forward limit, $Q^2$ is therefore the only semihard external scale entering
our rapidity prescription. \\

For the parametrization of the non-perturbative target correlator at the initial condition $\eta= \eta_0$, we use the ($\gamma$-modified) McLerran-Venugopalan model~\cite{McLerran:1993ni,McLerran:1994vd,McLerran:1993ka}
\begin{gather}
    \mathcal{U} (\mathbf{r} , \eta= \eta_0) \equiv \mathcal{U} (r, \eta= \eta_0) = 1 - \exp^{-\frac{(r^2 Q_0^2)^{\gamma}}{4} \ln \left( \frac{1}{ r \Lambda} + e \right)} \; ,
    \label{Eq:MV}
\end{gather}
with $\Lambda = 0.241 \; {\rm GeV}$, $e$ the Euler number and $r= |\mathbf{r}|$. $\gamma$ is a standard fit parameter employed in the literature. The parameter $Q_0^2$ is chosen as $Q_0^2 = 0.104 \; {\rm GeV}^2$ for a proton target, and as $Q_0^2 = 6 \times 0.104 \; {\rm GeV}^2$ for a large nucleus, where the factor $6$ corresponds to the geometric scaling factor $A^{1/3}$ for a lead target. \\

We implement the evolution of the dipole operator, including running-coupling effects, exactly as in~\cite{Cougoulic:2024jnd}, {\it i.e.}\ following the derivation of Ref.~\cite{Ducloue:2019ezk}. For completeness, we reproduce below the form of the equation implemented in our numerical analysis,
\begin{gather} 
\frac{\partial \mathcal{U} ( r, \eta  ) }{\partial \eta} = \int d \phi d r_z r_z\left[\frac{\bar{\alpha}_s(r)}{2 \pi r_z^2}\left(\frac{r^2}{r_{z y}^2+\epsilon^2}+\frac{\bar{\alpha}_s\left(r_z\right)}{\bar{\alpha}_s\left(r_{z y}\right)}-1+\frac{r_z^2}{r_{z y}^2+\epsilon^2}\left(\frac{\bar{\alpha}_s\left(r_{z y}\right)}{\bar{\alpha}_s\left(r_z\right)}-1\right)\right)\right] \nonumber \\ \times\left[ \mathcal{U} \left(r_z, \eta-\delta_{r_z ; r}\right) + \mathcal{U} \left(r_{z y}, \eta-\delta_{r_{z y} ; r}\right) - \mathcal{U} ( r, \eta ) - \xi \; \mathcal{U} \left(r_z, \eta-\delta_{r_z ; r}\right) \mathcal{U} \left(r_{z y}, \eta-\delta_{r_{z y} ; r}\right) \right] \; ,
\label{Eq:EvolEquation}
\end{gather}
where 
\begin{gather*}
    r_{zy} = \sqrt{r^2 + r_z^2 - 2 r r_z \cos \phi} \; ,  \\ \hspace{0.5 cm} \delta_{r_z;r} = \max \left \{ 0 , 2 \ln \left( \frac{r}{r_{z}} \right) \right \} \; , \hspace{0.5 cm} \delta_{r_{zy};r} = \max \left \{ 0 , 2 \ln \left( \frac{r}{r_{zy}} \right) \right \} \; .
\end{gather*}
The prescription for $\bar{\alpha}_s$ is the one introduced in Ref.~\cite{Lappi:2013zma} and quoted in eqs.~(5) and (6) of~\cite{Cougoulic:2024jnd}. We emphasize that the only difference with respect to the implementation of~\cite{Cougoulic:2024jnd} is that we write the evolution in terms of the dipole operator  $\mathcal{U}(r,\eta) = 1 - S(r,\eta)$, whereas they formulate it in terms of $S(r,\eta)$. Furthermore, we introduce a parameter $\xi$ that allows us to switch from the non-linear BK equation ($\xi = 1$) to the linear BFKL limit ($\xi = 0$). We validated our implementation by comparing the $\xi = 1$ case with the results of~\cite{Cougoulic:2024jnd}, finding agreement at the permille level. In the following, for simplicity of notation, we write  $\mathcal{U}(r,\eta) \equiv \mathcal{U}(r)$. It is important to clarify the scope of this comparison. The MV initial
condition in Eq.~(\ref{Eq:MV}) already incorporates multiple soft scatterings and is therefore not a purely dilute initial condition, even when it is subsequently
evolved with the linear equation, $\xi=0$. We deliberately use the same
functional form of the initial condition for both evolutions in order to minimize the dependence of the comparison on nonperturbative modeling and to
isolate the effect of the nonlinear term in the perturbative small-$x$
evolution. Thus, throughout this work, the labels BFKL and BK refer,
respectively, to linear and nonlinear evolution starting from the same
multiple-scattering initial condition. While this initial condition is
unitarized at the starting rapidity, linear BFKL evolution does not preserve
this property at sufficiently large rapidity. \\

To assess the dependence of our results on the evolution prescription, we also consider alternative implementations of the small-$x$ evolution, both with and without collinear improvement, as well as different parametrizations of the dipole initial condition. These alternatives will be compared in Sec.~\ref{sec:numerical} as part of our estimate of the dominant theoretical uncertainty.

\subsection{Twist-2 and -3 distribution amplitudes}
The additional non-perturbative ingredients required in our analysis are the explicit parame\-trizations of the higher-twist collinear distribution amplitudes (DAs). These were originally derived in Ref.~\cite{Ball:1998sk}. In addition to their asymptotic forms, we implement their renormalization-group evolution. The scale dependence of the DAs enters their expressions below through the coupling constants $a_2 (\mu^2)$, $\omega^A_{\{1,0\}} (\mu^2)$, $f_{3M}^A (\mu^2)$, and $f^V_{3M} (\mu^2)$, whose evolution equations are given in Ref.~\cite{Ball:1998sk}. For the convenience of the reader, Appendix~\ref{App:RenGroup} collects both the evolution equations and the values of these constants for the $\rho$ meson ($M = \rho$) at $\mu^2 = 1$ GeV$^2$, which are used in our numerical analysis. In this analysis, we take the $\rho$ meson mass to be $m_{\rho} = 0.776$ GeV and the vector decay constant $f_{\rho} = 200$ MeV~\cite{Becirevic:1998jp}.

\paragraph{Distribution amplitudes parametrization.} The longitudinal meson DA is parameterized as~\cite{Anikin:2009bf}
\begin{equation}
    \varphi_{\parallel} (z) = 6 z (1-z) \left( 1 + \frac{3}{2} a_2 \left( \mu^2 \right) (5 ( 2z - 1 )^2 -1) \right) \; .
    \label{Eq:IndipDALon}
\end{equation}
We denote by $z$ the fraction of the meson longitudinal ``$+$'' momentum
carried by the quark, $z=k_q^+/p_M^+$ , while the antiquark carries the fraction $\bar z=1-z$. In the
eikonal approximation, longitudinal momentum is conserved through the
scattering, so the same fractions refer also to the incoming photon, {\it i.e.} $p_M^+ = q^+$. 
\\

The genuine twist-3 DAs, in terms of which the results of Ref.~\cite{Boussarie:2024bdo} can be expressed, read~\cite{Anikin:2009bf}
\begin{equation}
    V(x_1, x_2) = 5040 \; x_1 x_2 (x_1 - x_2) (1-x_1 - x_2)^2  \; , 
    \label{Eq:IndipDAVect3}
\end{equation}
\begin{equation}
     A (x_1, x_2) = 360 x_1 x_2  \left( 1 + \frac{ \omega^A_{\{ 1,0\} } ( \mu^2 ) }{2} (4 - 7 (x_1 + x_2) ) \right) (1-x_1 - x_2)^2 \; ,
      \label{Eq:IndipDAAxial3}
\end{equation}
where, $x_1$, $x_2$, and $x_3$ denote the fractions of
the meson longitudinal ``$+$'' momentum carried by the quark ($x_1 = k_q^+/p_M^+$), antiquark ($x_2 = k_{\bar q}^+/p_M^+$), and
gluon ($x_3 = k_g^+/p_M^+$), respectively, with $x_1+x_2+x_3=1$. By eikonal longitudinal-momentum
conservation, these are equivalently the corresponding fractions of the
incoming photon momentum. \\

In Ref.~\cite{Boussarie:2024bdo}, the 2-body twist-3 corrections are expressed in terms of the DAs
\begin{equation}
       \varphi_1^T (z) \equiv \varphi_1^T (z) \big |_{ \rm W. W.} + \varphi_1^T (z) \big |_{ \rm G. T.} = - h(z) + \Tilde{h} (z) \; ,
       \label{Eq:Varphi1T}
\end{equation}
\begin{equation}
          \varphi_A^T (z) \equiv \varphi_A^T (z) \big |_{ \rm W. W.} + \varphi_A^T (z) \big |_{ \rm G. T.} = - \frac{ g_{\perp}^{(a)} (z) }{ 4 }  + \frac{\Tilde{g}_{\perp}^{(a)} (z) }{ 4 }  \; ,
           \label{Eq:VarphiAT}
\end{equation}
where W.W. (G.T.) denotes the Wandzura–Wilczek (genuine twist-3) component of the DA. The final equalities in both expressions are given for easier comparison with the formulas of Ref.~\cite{Boussarie:2024bdo}. \\

The four contributions in eqs.~(\ref{Eq:Varphi1T}) and~(\ref{Eq:VarphiAT}) can be expressed in terms of the independent set of DAs $\{\varphi_\parallel, V, A\}$, parametrized in eqs.~(\ref{Eq:IndipDALon}),~(\ref{Eq:IndipDAVect3}), and~(\ref{Eq:IndipDAAxial3}). The derivation of the explicit parameterizations of contributions in eqs.~(\ref{Eq:Varphi1T}) and~(\ref{Eq:VarphiAT}) is long, we relegate it to the Appendix~\ref{App:ExpParam}. We obtain
\begin{gather}
    \varphi_1^T (z) \big |_{ \rm W. W.} 
    = \frac{3}{2} z \left(2 z^2-3 z+1\right)
   \bigg( 1 + a_2 \left( \mu^2 \right) ( 1 + 15 z (z-1) ) \bigg) \; ,
\end{gather}
\begin{gather}
\varphi_1^T (z) \big |_{ \rm G. T.} = -\frac{5}{8} z \left(2 z^2-3 z+1\right) \bigg[  \left(3 \omega^A_{\{ 1,0\} } ( \mu^2 ) \left(7 z^2-7 z+1\right)-16\right) \frac{f_{3M}^A}{f_{M}} \nonumber \\ -84 \left(23 z^2-23 z+1\right) \frac{f_{3M}^V}{f_{M}} \bigg] \; ,
\end{gather}
\begin{gather}
    \varphi_A^T (z) \big |_{ \rm W. W.} = \frac{3}{2} z (z-1) \bigg( 1+ a_2 \left( \mu^2 \right) (1+ 5 (z-1) z) \bigg) \; ,
\end{gather}
\begin{gather}
   \varphi_A^T (z) \big |_{ \rm G. T.} = \frac{5}{8} z (1-z) \bigg[ \left( \omega^A_{\{ 1,0\} } ( \mu^2 )  \left(63 z^2-63 z+3\right) -16 \left(14 z^2-14 z+1\right) \right) \frac{f_{3M}^A}{f_{M}} \nonumber \\  -84 \left(5 z^2-5 z+1\right) \frac{f_{3M}^V}{f_{M}} \bigg] \; .
\end{gather}

\section{Helicity amplitudes}
In this section, we present the final analytic form of all helicity amplitudes of the unpolarized DVMP. We distinguish contributions arising from the quark-antiquark (2-body) Fock state, which dominates at leading twist, from those involving the quark-antiquark-gluon (3-body) configurations, relevant at twist-3 accuracy. For the sake of generality, we will consider the matrix element $\mathcal{U} (\mathbf{r})$ in full generality in this section, although we will use the parametrization~(\ref{Eq:MV}) in the numerical analysis. In deriving the analytic expressions below, we keep the dipole matrix element
$U(\boldsymbol r,\eta)$ general, assuming azimuthal symmetry only where
explicitly stated in the simplification of the 3-body contribution. The
numerical analysis instead employs the dipole operator obtained by evolving
the MV initial condition in Eq.~(\ref{Eq:MV}) according to Eq.~(\ref{Eq:EvolEquation}).

\subsection{Helicity amplitudes: 2-body Fock state}
When only the leading quark-antiquark Fock state is taken into account, the small-$x$ amplitude can be written in the dipole form
\begin{equation}
    \mathcal{A}_2^{\lambda_{V} , \lambda_{\gamma} }  = (2 \pi)^4 \delta^{(4)} (q + p_t - p_M -p_{t'} ) \; W^2 \int d^2 \boldsymbol{r}  \; \mathcal{U} \left( \boldsymbol{r} \right) \int_0^1 d z \; e^{i z\boldsymbol{\Delta} \cdot \boldsymbol{r} } \; \psi_2^{  \lambda_{V}  , \lambda_{\gamma}} \left( z, \boldsymbol{r} \right)  ,
    \label{Eq:Two-body_dipole_amp}
\end{equation}
where $\lambda_{\gamma} = \{T,L \} $ and $\lambda_{V} = \{T,L \}$ represent the incoming photon and outgoing meson helicities, respectively. In the frame in which the incoming photon has no transverse momentum in the Sudakov sense, the coordinate-space impact factors for the different helicity amplitudes read~\cite{Boussarie:2016bkq,Boussarie:2024bdo} 
\begin{equation}
    \psi_2^{L , L } \left( z, \boldsymbol{r} \right) = \frac{e_q m_M f_M}{2 \pi} 2 z \bar{z} Q^2 K_0 \left( \sqrt{z \bar{z} Q^2 \boldsymbol{r}^2} \right) \left[ \frac{\varepsilon_{\gamma L}^+}{q^+} \frac{\varepsilon_{ML}^{+*}}{p_M^+} \right] \varphi_{\parallel} (z) \; ,
\end{equation}
\begin{equation}
    \psi_2^{L , T } \left( z, \boldsymbol{r} \right) = \frac{e_q m_M f_M }{2 \pi} (2z-1) \sqrt{ \frac{z \bar{z} Q^2}{\boldsymbol{r}^2}} K_1 \left( \sqrt{z \bar{z} Q^2 \boldsymbol{r}^2} \right) \left[ i \frac{\varepsilon_{ML}^{+*}}{p_M^+} (\boldsymbol{r} \cdot \boldsymbol{\varepsilon}_{\gamma} ) \right] \varphi_{\parallel} (z) \; ,
\end{equation}
\begin{equation}
    \psi_2^{ T, L } \left( z, \boldsymbol{r} \right) = \frac{e_q m_M f_M }{2 \pi} 2 z \bar{z} Q^2 K_0 \left( \sqrt{z \bar{z} Q^2 \boldsymbol{r}^2} \right) \left[ i \frac{\varepsilon_{\gamma L}^+}{q^+} (\boldsymbol{r} \cdot \boldsymbol{\varepsilon}_M^{*} ) \right] \varphi_1^T (z) \; ,
\end{equation}
\begin{gather}
    \psi_2^{T , T } \left( z, \boldsymbol{r} \right) = - \frac{e_q m_M f_M }{2 \pi} \sqrt{z \bar{z} Q^2 \boldsymbol{r}^2} K_1 \left( \sqrt{z \bar{z} Q^2 \boldsymbol{r}^2} \right) \left[  \frac{T_{\rm n. f.}}{2} \; \phi_{2, { \rm n. f.}} (z) + T_{\rm f.} (\boldsymbol{r}) \; \phi_{2, {\rm f.}} (z)  \right] \nonumber \\ \equiv \psi_{2, {\rm n. f.}}^{T , T } + \psi_{2, {\rm f.}}^{T , T } \;,
\end{gather}
where we introduced the flip (f.) and non-flip (n.f.) helicity structures
\begin{equation}
   T_{\rm f.} (\boldsymbol{r}) = \frac{ \left( \boldsymbol{\varepsilon}_{\gamma} \cdot \boldsymbol{r} \right) \left( \boldsymbol{\varepsilon}_{M}^{*} \cdot \boldsymbol{r} \right) }{\boldsymbol{r}^2} - \frac{\boldsymbol{\varepsilon}_{\gamma} \cdot \boldsymbol{\varepsilon}_{M}^{*} }{2} \; , \hspace{1 cm} T_{\rm n. f.} = \boldsymbol{\varepsilon}_{\gamma} \cdot \boldsymbol{\varepsilon}_{M}^{*} \; ,
\end{equation}
and the combination of DAs
\begin{equation}
    \phi_{2, {\rm f.}} (z) = -(2 z - 1) \varphi_1^T (z) +  \varphi_A^T (z) \;,
\end{equation}
\begin{equation}
    \phi_{2, {\rm n. f.}} (z) = - (2 z - 1) \varphi_1^T (z)  -  \varphi_A^T (z) \;.
\end{equation}
At the level of Wandzura-Wilczek approximation, it is enough to take the W.W. part of the DAs in eqs.~(\ref{Eq:Varphi1T},  \ref{Eq:VarphiAT}), while in a fully consistent twist-3 treatment, also the G.T. part must be included. The helicity structures are such that
\begin{equation}
    T_{\rm f.}^{(+) (+)} = T_{\rm f.}^{(-) (-)} = 0  \; , \hspace{0.5 cm}   T_{\rm f.}^{(-) (+)} = \left( T_{\rm f.}^{(+) (-)} \right)^{*} = - \frac{1}{2} e^{2 i \varphi} \; ,
\end{equation}
\begin{equation}
    T_{\rm n. f.}^{(+) (+)} = T_{\rm n. f.}^{(-) (-)} = 1  \; , \hspace{0.5 cm}   T_{\rm n. f.}^{(-) (+)} = \left( T_{\rm n. f.}^{(+) (-)} \right)^{*} = 0  \; ,
\end{equation}
where $\varphi$ is is the polar angle of $\boldsymbol{r}$ with respect to the $x$ axis. \\

Let us analyze the scaling behaviour of the different helicity amplitudes. In the following, we will use $\lambda_{\gamma} \; ( {\rm or} \; \lambda_{V}) = 0$ for longitudinal polarization,  $\lambda_{\gamma} \; ( {\rm or} \; \lambda_{V}) = 1$ for ``plus'' ($\varepsilon^{(+)}$) transverse polarization and $\lambda_{\gamma} \; ( {\rm or} \; \lambda_{V}) = -1$ for ``minus'' ($\varepsilon^{(-)}$) transverse polarization. The longitudinal-photon-to-longitudinal-meson amplitude is non-zero in the forward limit and scales for $Q \rightarrow \infty$ as
\begin{gather}
    \mathcal{A}_2^{0,0} \sim Q \int d r \; r \;  K_0 \left( \sqrt{z \bar{z} Q^2 r^2} \right) \sim \frac{1}{Q} \; ,
\end{gather}
thus being leading twist. The transverse-photon-to-transverse-meson $s$-channel helicity-conserving (non-flip) amplitude is non-zero in the forward limit and scales for $Q \rightarrow \infty$ as
\begin{gather}
    \mathcal{A}_2^{1,1} \sim Q \; m_M \int d r \; r^2 \; K_1 \left( \sqrt{z \bar{z} Q^2 r^2} \right) \sim \frac{1}{Q} \frac{m_M}{Q} \; .
\end{gather}
The $s$-channel non-conserving helicity amplitudes are zero in the forward limit. For the longitudinal (transverse)-photon-to-transverse (longitudinal)-meson amplitude, $\mathcal{A}_2^{0,1}$ ($\mathcal{A}_2^{1,0}$), if we set $|\boldsymbol{\Delta}|=0$ in the phase factor in eq.~(\ref{Eq:Two-body_dipole_amp}), then the integral over $z$ vanishes due to the anti-symmetric (symmetric) property of the $\varphi_{1}^T (z)$ ($\varphi_{ \parallel} (z)$) under the exchange $z \rightarrow 1-z$. The first leading contribution in the next-to-forward kinematics comes from the expansion at the first order in $|\boldsymbol{\Delta}|$ of the phase factor and leads to the scaling behavior 
\begin{gather}
    \mathcal{A}_2^{0,1} \sim Q \; |\boldsymbol{\Delta}| \int d r \; r^2 \; K_1 \left( \sqrt{z \bar{z} Q^2 r^2} \right) \sim \frac{1}{Q} \frac{|\boldsymbol{\Delta}|}{Q} \; ,
\end{gather}
\begin{gather}
    \mathcal{A}_2^{1,0} \sim Q \; m_M \; |\boldsymbol{\Delta}| \int d r \; r^3 \; K_0 \left( \sqrt{z \bar{z} Q^2 r^2} \right) \sim \frac{1}{Q} \frac{m_M}{Q} \frac{|\boldsymbol{\Delta}|}{Q} \; ,
\end{gather}
in the $Q \rightarrow \infty$ limit. The transverse-photon-to-transverse-meson $s$-channel helicity-non-conserving (flip) amplitude is particularly interesting, because it is zero in the forward and next-to-forward limits. Indeed, in this case, the vanishing is due to the angular integrals 
\begin{gather}
    \int_0^{2 \pi} d \varphi \; e^{\pm i 2 \varphi } = 0 \; \; \; \rm and \;  \; \int_0^{2 \pi} d \varphi \; \cos \varphi \; e^{\pm i 2 \varphi } = 0\; ,
\end{gather}
appearing at forward and next-to-forward accuracy, respectively. If one goes beyond that in the perturbative part of the amplitude, the scaling reads 
\begin{gather}
    \mathcal{A}_2^{-1,1} \sim Q \; m_M \; |\boldsymbol{\Delta}|^2 \int d r \; r^4 \; K_1 \left( \sqrt{z \bar{z} Q^2 \boldsymbol{r}^2} \right) \sim \frac{1}{Q} \frac{m_M}{Q} \frac{|\boldsymbol{\Delta}|^2}{Q^2}  \; .
\end{gather}
Thus, if $m_{M} \gg |\boldsymbol{\Delta}|$, one find the hierarchy $ \mathcal{A}_2^{0,0} > \mathcal{A}_2^{1,1} > \mathcal{A}_2^{0,1} > \mathcal{A}_2^{1,0} , \mathcal{A}_2^{-1,1}$, experimentally observed at HERA~\cite{Aaron:2009xp}. 

\subsection{Helicity amplitudes: 3-body Fock state}
At twist-3 accuracy, the amplitude receives contributions from the three-body Fock state, which includes the quark, the antiquark, and an additional gluon. The most general structure of this contribution can be written in terms of the overlap between the photon/meson wavefunctions and Wilson line correlators involving up to four Wilson lines. This expression involves an integration over the longitudinal momentum fractions of the constituents, constrained by momentum conservation, as well as over their transverse positions. In Ref.~\cite{Boussarie:2024bdo}, it was found to be
\begin{gather}
    \mathcal{A}_3 = \left( \prod_{i=1}^3 \int d x_i \theta (x_i) \right) \delta \left( 1 - \sum_i^3 x_i \right) \int d^2 \boldsymbol{z}_1 d^2 \boldsymbol{z}_2 d^2 \boldsymbol{z}_3 e^{i \boldsymbol{q} (x_1 \boldsymbol{z}_1 + x_2 \boldsymbol{z}_2 + x_3 \boldsymbol{z}_3)} \nonumber \\
   \times \Psi_3 \left( \{ x \} , \{ \boldsymbol{z} \} \right) \left\langle P\left(p^{\prime}\right)\left| \mathcal{U}_{\boldsymbol{z}_1 \boldsymbol{z}_3} \mathcal{U}_{\boldsymbol{z}_3 \boldsymbol{z}_2} - \mathcal{U}_{\boldsymbol{z}_1 \boldsymbol{z}_3} -\mathcal{U}_{\boldsymbol{z}_3 \boldsymbol{z}_2} + \frac{1}{N_c^2} \mathcal{U}_{\boldsymbol{z}_1 \boldsymbol{z}_2}  \right|P\left(p\right)\right\rangle \; , \label{Eq:GenealStructure3bodyWaveFunOver}
\end{gather}
where the photon/meson wavefunction overlap reads
\begin{gather}
    \Psi_3 \hspace{-0.1 cm} \left( \{ x \} , \{ \boldsymbol{z} \} \right) \hspace{-0.1 cm} = \hspace{-0.1 cm} \frac{e_{q} m_M }{8 \pi}  \hspace{-0.1 cm} \left( \hspace{-0.1 cm} \varepsilon_{q\rho}-\frac{\varepsilon_{q}^{+}}{q^{+}}q_{\rho} \hspace{-0.1 cm} \right)  \hspace{-0.1 cm} \left( \hspace{-0.1 cm} \varepsilon^{* \mu }_{M} - \frac{p_{M}^{\mu}}{p_M^+} \varepsilon_{M}^{*+} \hspace{-0.1 cm} \right) \hspace{-0.1 cm} \delta \hspace{-0.1 cm} \left( \hspace{-0.1 cm} 1 - \frac{p_M^+}{q^+} \hspace{-0.1 cm} \right) \hspace{-0.15 cm} \left( \prod_{j=1}^{3} \theta (x_j) \theta (1-x_j) e^{-i x_j \boldsymbol{p}_M \boldsymbol{z}_j } \hspace{-0.1 cm} \right)   \nonumber \\
    \times c_f \bigg \{ f_{3M}^{V} g_{\sigma \mu} V(x_1, x_2) \left[ \left( 4 g_{\perp}^{ \rho \sigma } \frac{x_1 x_2}{1-x_2} \frac{Q}{Z} K_1 (QZ)  - i T_1^{\sigma \rho \nu} (\{ x \}) \frac{z_{23 \perp \nu}}{\boldsymbol{z}_{23}^{2}} K_0 (QZ) \right) - \left( 1 \leftrightarrow 2 \right) \right] \nonumber \\ 
    - \epsilon_{- + \sigma \beta} f_{3M}^{A} g^{\beta}_{\perp \mu} A (x_1, x_2) \hspace{-0.1 cm} \left[ \left( \hspace{-0.1 cm} 4 \epsilon^{ \sigma \rho + - } \hspace{-0.05 cm} \frac{x_1 x_2}{1-x_2} \frac{Q}{Z} K_1 (QZ) \hspace{-0.1 cm} + \hspace{-0.1 cm} T_2^{\sigma \rho \nu} (\{ x \}) \frac{z_{23 \perp \nu}}{\boldsymbol{z}_{23}^{2}} K_0 (QZ) \hspace{-0.1 cm} \right) \hspace{-0.1 cm} + \hspace{-0.1 cm} \left( 1 \leftrightarrow 2 \right) \hspace{-0.05 cm} \right] \hspace{-0.1 cm} \bigg \} ,
\end{gather}
with 
\begin{equation}
\begin{split}
   & T_1^{\sigma \rho \nu} ( \{ x \} )  \\
   &  = 4\left[2x_{1}q^{\rho}\left(2\frac{x_{2}}{x_{3}}+1\right)g_{\perp}^{\sigma\nu}-\left(2\frac{x_{1}-x_{2}}{x_{3}}g_{\perp}^{\sigma\nu}g_{\perp}^{\rho \alpha}-\frac{g_{\perp}^{\alpha \nu}g_{\perp}^{\rho\sigma}+g_{\perp}^{\sigma\nu}g_{\perp}^{\rho \alpha}-g_{\perp}^{\alpha \sigma}g_{\perp}^{\rho\nu}}{x_{2}+x_{3}}\right)i \partial_{z_{1\perp} \alpha} \right] \nonumber \\ & \equiv 4 \left[ 2 x_{1}q^{\rho}\left(2\frac{x_{2}}{x_{3}}+1\right) g_{\perp}^{\sigma\nu} - \Xi_1^{\alpha \nu \rho \sigma} i \partial_{z_{1\perp} \alpha} \right] ,
\end{split}
\end{equation} 
\begin{equation}
\begin{split}
   & T_2^{\sigma \rho \nu} ( \{ x \} ) \\
   & \hspace{-0.2 cm} = \frac{ 4 i }{ \bar{x}_1 } \left[ 2 x_1 \bar{x}_1 q^{\rho} \epsilon^{ \nu \sigma + -} - \left( \left( 1 + \frac{2 x_2}{x_3} \right) \left( g_{\perp}^{ \sigma \alpha } \epsilon^{\nu \rho + -} -  g_{\perp}^{ \rho \sigma } \epsilon^{\nu \alpha + -}  \right)  + (x_1 - \bar{x}_1) g_{\perp}^{ \rho \alpha }  \epsilon^{\nu \sigma + -} \right ) i \partial_{z_{1\perp} \alpha}  \right ]  \nonumber \\ & \equiv \frac{ 4 i }{ \bar{x}_1 } \bigg[ 2 x_1 \bar{x}_1 q^{\rho} \epsilon^{ \nu \sigma + -} - \Xi_2^{\alpha \nu \rho \sigma} i \partial_{z_{1\perp} \alpha} \bigg] \; ,
\end{split}
\end{equation}
and
\begin{equation}
 c_f = \frac{N_c^2}{N_c^2-1} \; , \hspace{1.5 cm}  Z = \sqrt{x_1 x_2 \boldsymbol{z}_{12}^2 + x_1 x_3 \boldsymbol{z}_{13}^2 + x_2 x_3 \boldsymbol{z}_{23}^2 } \; \; .
\end{equation}

The $(1 \leftrightarrow 2)$ transformation means $(x_1, \boldsymbol{z}_{1} \leftrightarrow x_2, \boldsymbol{z}_{2})$ and should not be confused with the $(x_1 \leftrightarrow x_2)$ transformation. \\

The completely general expression given above can be strongly simplified in the pure dipole limit, {\it i.e.} neglecting the product of two dipole operators in eq.~(\ref{Eq:GenealStructure3bodyWaveFunOver})\footnote{Please note that this does not mean neglecting non-linear saturation effects due to the evolution, which are incorporated through the evolution of the dipole.}. In this limit, the 6-dimensional integration over the transverse coordinates can be reduced to a 2-dimensional integral, significantly easing numerical evaluations. Furthermore, in models with radial symmetry, such as the McLerran–Venugopalan model, angular integrations become trivial. The detailed derivation of these simplifications is provided in Appendix~\ref{App:DerivAmp}, where both the main steps and the required integrals are worked out. \\

In the numerical analysis of the present work, we will consider only the $T \rightarrow T$ transition, which is the only one surviving in the forward limit. However, by performing a non-trivial expansion around $\Delta=0$ ({\it i.e.} relaxing the strictly forward direction), we will also extract, for the first time, the dominant contribution of the longitudinal-photon-to-transverse-meson amplitudes in the quasi-forward limit. \\

\noindent To streamline the notation in the results that follow, we define effective mass parameters associated with the longitudinal momentum fractions,
\begin{equation}
   \mu_i = \sqrt{x_i \bar{x}_i Q^2} \; ,  \hspace{0.5 cm} \mu_{ij} = \sqrt{ \frac{x_i x_j}{x_i + x_j} Q^2}\;,
\end{equation}
as well as the dimensionless ratios of the twist-3 decay constants to the leading-twist one,
\begin{equation}
    \zeta_{3M}^A = \frac{f_{3M}^A}{f_M} \; , \hspace{0.5 cm} \zeta_{3M}^V = \frac{f_{3M}^V}{f_M} \;.
    \label{Eq:Zeta}
\end{equation}
Moreover, we define the overall prefactors
\begin{gather}
    \mathcal{C}_{L} = - (2 \pi)^4 \delta^{(4)} (q + p_t - p_M -p_{t'} ) W^2 \; e_q m_M f_M \; , \nonumber \vspace{0.1 cm} \\ \mathcal{C}_T = - 2 (2 \pi)^4 \delta^{(4)} (q + p_t - p_M -p_{t'} ) W^2 \; \frac{e_q m_M f_M}{Q^2}  \; .
\end{gather}
encoding the normalization of the longitudinal and transverse amplitudes, respectively.

\paragraph{Longitudinal-photon-to-transverse-meson amplitude.} Starting from eq.~(\ref{Eq:A-LT-AlmostFin}) and replacing the explicit expression of the $F_i (x_i,x_j,\boldsymbol{r}^2)$ and $\tilde{F}_i (x_i,x_j,\boldsymbol{r}^2)$ functions in the appendix~\ref{App:F_i_functions}, for the longitudinal-to-transverse helicity transition, we obtain
\begin{gather}
    \mathcal{A}^{T,L}_{3} =  \mathcal{C}_L \int \frac{d^2 \boldsymbol{r}}{2 \pi} \; \mathcal{U} (\boldsymbol{r}) \; \left[ (\boldsymbol{\varepsilon}^{*}_M \cdot \boldsymbol{\Delta} ) \frac{\varepsilon_q^+}{q^+} \right] \left[ \prod_{i=1}^3 \int d x_i \right]  \delta \left(1-\sum_i^3 x_i \right) \nonumber \\ \times \left( \zeta_{3M}^A A(x_1, x_2) - \left( 1 + \frac{  2 x_2  }{x_3} \right) \zeta_{3M}^V V (x_1, x_2) \right)  \nonumber \\ \times \left \{  \left( \frac{N_c}{C_F} - 2 \right) \left[ - \frac{x_1}{x_3}  \mu_1 |\boldsymbol{r}| K_1 \left( \mu_1 |\boldsymbol{r}| \right) + \frac{ x_1 }{x_3 (x_1+x_2)} \mu_{12} |\boldsymbol{r}| K_1 \left( \mu_{12} |\boldsymbol{r}| \right) \right] \right. \nonumber \\ \left. + \frac{N_c}{C_F} \left[ - \frac{x_1}{x_2} \mu_1 |\boldsymbol{r}| K_1 (\mu_1 |\boldsymbol{r}|) + \frac{x_1}{x_2 (x_1+x_3) } \mu_{13} |\boldsymbol{r}| K_1 (\mu_{13} |\boldsymbol{r}|)  + \frac{x_2}{(x_2+x_3)^2} \mu_{23} |\boldsymbol{r}| K_1 (\mu_{23} |\boldsymbol{r}|) \right] \right \} .
\end{gather}

\paragraph{Transverse-photon-to-transverse-meson-transition amplitudes.}
Similarly, the helicity non-flip part of the transverse-to-transverse helicity amplitude reads
\begin{gather}
    \mathcal{A}^{T,T}_{3, \rm n. f.} =  \mathcal{C}_T \int \frac{d^2 \boldsymbol{r}}{2 \pi} \; \mathcal{U} (\boldsymbol{r}) \; \frac{T_{\rm n. f.}}{2} \left[ \prod_{i=1}^3 \int_0^1 d x_i \right] \delta \left(1-\sum_i^3 x_i \right) \left( \frac{ \zeta_{3M}^A A(x_1, x_2) + \zeta_{3M}^V V (x_1, x_2) }{x_3} \right)  \nonumber \\ \times \left \{  \left( \frac{N_c}{C_F} - 2 \right) \left[ \frac{1}{x_3}  \mu_1^2 K_0 \left( \mu_1 |\boldsymbol{r}| \right) + \frac{\bar{x}_2}{x_2 x_3} \mu_2^2 K_0 \left( \mu_2 |\boldsymbol{r}| \right) - \frac{1}{x_3} \left( 1 + \frac{\bar{x}_2}{x_2} \right) \mu_{12}^2 K_0 \left( \mu_{12} |\boldsymbol{r}| \right) \right] \right. \nonumber \\ \left. + \frac{N_c}{C_F} \left[ \frac{\bar{x}_2}{x_1 x_2} \mu_2^2 K_0 (\mu_2 |\boldsymbol{r}|) + \frac{1}{\bar{x}_1} \left( 1 - \frac{\bar{x}_1 \bar{x}_2}{x_1 x_2} \right) \mu_{23}^2 K_0 (\mu_{23} |\boldsymbol{r}|) \right] + \frac{2}{\bar{x}_1} \mu_1^2 K_0 (\mu_1 |\boldsymbol{r}|) \right \} \; ,
\end{gather}
while the helicity flip part is
\begin{gather}
    \mathcal{A}^{T,T}_{3, \rm f.} =  \mathcal{C}_T \int \frac{d^2 \boldsymbol{r}}{2 \pi} \; \mathcal{U} (\boldsymbol{r}) \; T_{\rm  f.}(\boldsymbol{r}) \left[ \prod_{i=1}^3 \int_0^1 d x_i \right] \frac{ \delta (1-\sum_i^3 x_i )}{x_1} \left( \frac{ \zeta_{3M}^A A(x_1, x_2) + \zeta_{3M}^V V (x_1, x_2) }{x_3} \right)  \nonumber \\ \times \bigg \{ \frac{N_c}{C_F} \bigg[ \mu_{qg}^2 \; K_2 (\mu_{qg} |\boldsymbol{r}| ) + \mu_{\bar{q}g}^2 \; K_2 (\mu_{\bar{q}g} |\boldsymbol{r}| ) - \mu_1^2 K_2 (\mu_1 |\boldsymbol{r}|) - \mu_2^2 K_2 (\mu_2 |\boldsymbol{r}|) \bigg] + \left( \frac{N_c}{C_F} - 2 \right) \nonumber \\ \times \bigg[ \frac{x_2}{x_3} \bigg( \mu_{q \bar{q}}^2 K_2 (\mu_{q \bar{q}}|\boldsymbol{r}|) - \mu_{1}^2 K_2 (\mu_{1}|\boldsymbol{r}|) \bigg) + \frac{x_1}{x_3} \bigg( \mu_{q \bar{q}}^2 K_2 (\mu_{q \bar{q}}|\boldsymbol{r}|) - \mu_{2}^2 K_2 (\mu_{2}|\boldsymbol{r}|) \bigg) \bigg] \bigg \} .
\end{gather}

\section{Numerical results}
\label{sec:numerical}

In this section, we first compare our framework with HERA measurements of exclusive vector meson electroproduction, focusing on two helicity-sensitive observables. We then turn to predictions for the future Electron–Ion Collider (EIC), where scattering off a heavy nuclear target (we consider lead) is expected to enhance saturation effects compared to the proton case. In this section, we present our phenomenological analysis of exclusive $\rho$-meson electroproduction. We first compare the helicity-sensitive observables with HERA data and investigate the theoretical uncertainty associated with the correlated choice of the dipole initial condition and small-$x$ evolution prescription. Despite a sizeable quantitative uncertainty, the resulting predictions generally reproduce the $Q^2$ dependence of the data, but tend to overestimate the magnitude of the observables. A consistent resolution would require a dedicated refit of the dipole initial condition to exclusive data, which is beyond the scope of this paper and is left for future work. For the present exploratory analysis, we instead adopt an effective phenomenological prescription based on the rapidity-scale parameter $k$. After tuning this latter, we examine the relative impact of linear and non-linear evolution and of genuine twist-3 contributions. Finally, we present predictions for the future Electron--Ion Collider (EIC), where scattering off a heavy nuclear target---here taken to be lead---is expected to enhance non-linear effects relative to the proton case.  \\

We consider two observables that were extensively studied at HERA in exclusive vector meson production~\cite{Aaron:2009xp}. The first one is the helicity-conserving transverse-to-longitudinal amplitude ratio
\begin{equation}
R_{11} = \frac{\mathcal{A}^{11}}{\mathcal{A}^{00}} \, ,
\end{equation}
This ratio provides direct information on the relative importance of transverse and longitudinal photon polarizations. The second observable is the spin-density matrix element $r_{00}^{04}$, which can be expressed in terms of helicity amplitudes as
\begin{equation}
r_{00}^{04} = \frac{\varepsilon + R_{01}^2}{R_{11}^2 + \varepsilon + R_{01}^2 + R_{1 -1}^2 + 2 \varepsilon R_{10}^2 } \,, 
\end{equation}
where $\varepsilon$ denotes the virtual photon polarization parameter and $R_{\lambda_V \lambda_\gamma} \equiv \mathcal{A}^{\lambda_V \lambda_\gamma} / \mathcal{A}^{00}$ are ratios of helicity amplitudes normalized to the longitudinal one.

In the limit of small momentum transfer ($t \to 0$), where helicity is approximately conserved (s-channel helicity conservation, SCHC), the helicity-flip amplitudes are suppressed, {\it i.e.} $R_{10}, R_{01}, R_{1 -1} \ll 1$. Under this assumption, the expression simplifies to
\begin{equation}
r_{00}^{04} \simeq \frac{\varepsilon}{R_{11}^2 + \varepsilon} \,.
\end{equation}
In the present analysis, we adopt this approximation, which is justified in the near-forward kinematics. However, it is important to stress that, beyond the strict SCHC limit, a direct comparison between $r_{00}^{04}$ and $R_{11}$ is non-trivial. While $R_{11}$ is defined purely in terms of helicity-conserving amplitudes, the experimental extraction of $r_{00}^{04}$ receives contributions from helicity-flip transitions and from non-zero momentum transfer. As a consequence, the one-to-one correspondence implied by the above relation is not exact at the experimental level, and the two observables should be regarded as complementary rather than redundant probes of the underlying dynamics. \\

For H1, one has $\langle \varepsilon \rangle = 0.98$, while for ZEUS $\langle \varepsilon \rangle = 0.996$, indicating an almost purely longitudinal photon flux. For the EIC, we will assume $\langle \varepsilon \rangle = 1$, which is an excellent approximation in the high-energy limit. 

\subsection{Theoretical uncertainties from the initial condition and small-$x$ evolution}
\label{sec:theoretical_uncertainties}

The dipole amplitude entering the helicity amplitudes is not determined by
perturbative small-$x$ evolution alone. Its phenomenological implementation
requires both a nonperturbative initial condition and a prescription for its
rapidity evolution. The parameters defining these two ingredients are strongly
correlated, since the initial-condition parameters are extracted from data
using a specific evolution kernel, running-coupling prescription, and starting
point. Consequently, the uncertainty associated with the dipole amplitude
cannot be reliably assessed by varying any of these ingredients in isolation.  

\begin{figure}[t]
    \centering
    \includegraphics[width=0.72\textwidth]
    {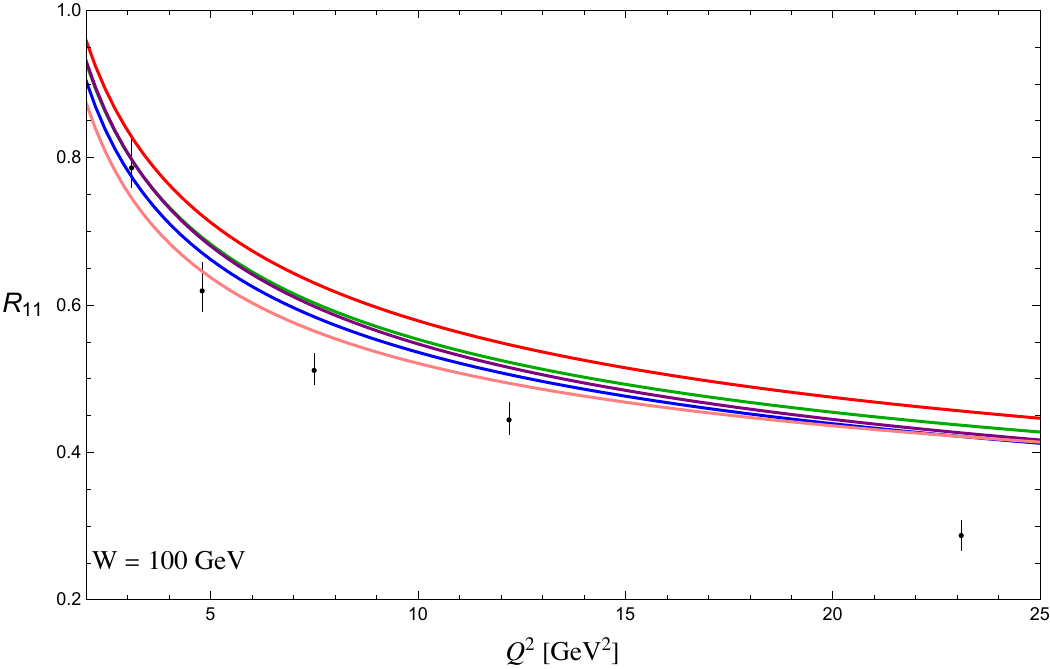}
    \caption{Predictions for the helicity-amplitude ratio $R_{11}$ obtained
    from the prescriptions summarized in
    Table~\ref{tab:dipole_prescriptions} versus HERA data. The colors identify the prescriptions summarized in Table~1.}
    \label{fig:R11_uncertainties}
\end{figure}

To investigate this dependence, in Fig.~\ref{fig:R11_uncertainties}
we compare predictions obtained from several representative combinations
of initial conditions and evolution prescriptions. These combinations allow
us to assess the sensitivity to the initial condition, the
running-coupling prescription, and the inclusion of collinear improvement.

\begin{table}[t]
    \centering
    \small
    \begin{tabular}{
        c
        p{0.34\textwidth}
        p{0.45\textwidth}
    }
        \hline
        Curve & Evolution equation & Initial condition from incusive DIS fit \\
        \hline
        Red &
        rcBK with collinear improvement as in
        Eq.~(\ref{Eq:EvolEquation})
        &
        HERA--BAL+SD with
        $\eta_0=0$, from Table~III of
        Ref.~\cite{Beuf:2020dxl} (MV$^\gamma$ initial condition)
        \vspace{0.2 cm} \\ 

        Green &
        rcBK with collinear improvement as in
        Eqs.~(27),~(32) of Ref.~\cite{Beuf:2020dxl}
        &
        HERA--BAL+SD with
        $\eta_0=0$, from Table~III of
        Ref.~\cite{Beuf:2020dxl} (MV$^\gamma$ initial condition)
        \vspace{0.2cm} \\

        Purple &
        rcBK with collinear improvement as in
        Eq.~(\ref{Eq:EvolEquation})
        &
        MV with $\eta_0=\ln(1/0.01)$, from Table~I of
        Ref.~\cite{Lappi:2013zma} (MV initial condition)
        \vspace{0.2cm} \\

        Blue &
        rcBK without collinear improvement, obtained from
        Eq.~(\ref{Eq:EvolEquation})
        &
        MV with $\eta_0=\ln(1/0.01)$, from Table~I of
        Ref.~\cite{Lappi:2013zma} (MV initial condition)
        \vspace{0.2cm} \\

        Pink &
        rcBK without collinear improvement, obtained from
        Eq.~(\ref{Eq:EvolEquation}) 
        &
        MV$^\gamma$ with $\eta_0=\ln(1/0.01)$, from Table~I of
        Ref.~\cite{Lappi:2013zma} (MV$^\gamma$ initial condition)
        \\
        
        \hline
    \end{tabular}
    \caption{Initial conditions and evolution prescriptions used in
    Fig.~\ref{fig:R11_uncertainties}. Here, BAL+SD denotes the Balitsky plus smallest-dipole running-coupling prescription.}
    \label{tab:dipole_prescriptions}
\end{table} 

Despite the sizeable quantitative spread, all these prescriptions produce
a similar decreasing $Q^2$ dependence, broadly following the trend of the
HERA data, while systematically overestimating the magnitude of $R_{11}$. The discrepancy is therefore not removed by changing the initial
condition, switching the collinear improvement on or off, or adopting a
different running-coupling prescription within the considered inclusive-DIS
fits. This indicates that the main limitation is the direct transfer of
dipole parametrizations constrained by inclusive DIS to the present exclusive
observable.

\begin{figure}[t]
    \centering    \includegraphics[width=0.65\textwidth]{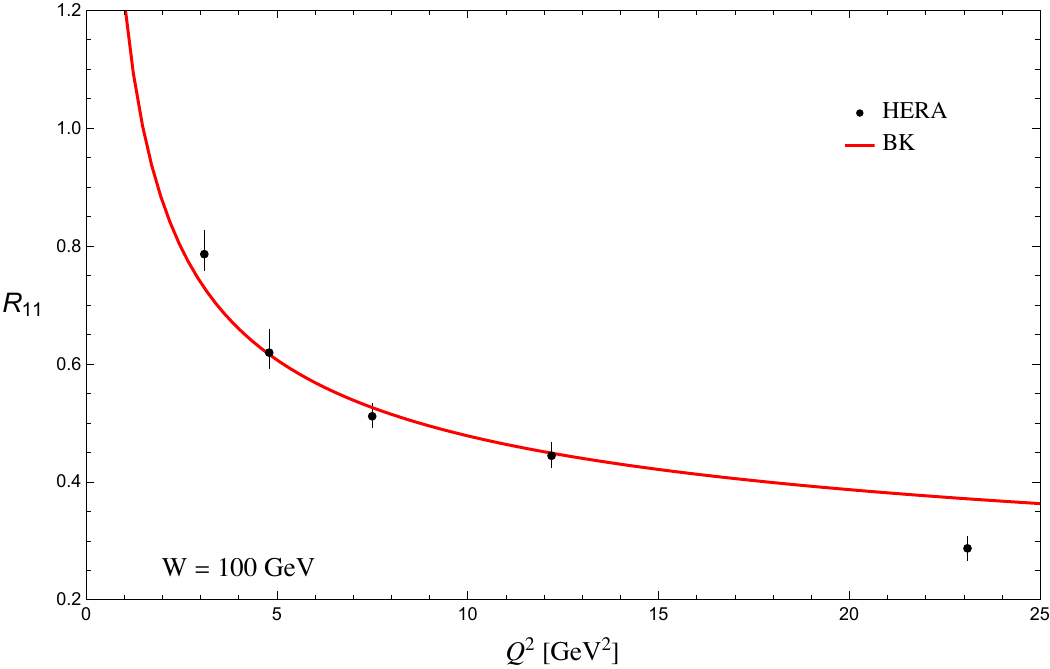}
    \caption{Prediction for $R_{11}$ obtained using the MV$^{\gamma}$ initial condition
    and running-coupling evolution of Ref.~\cite{Lappi:2013zma}, with
    $x_0=0.01$, no additional collinear improvement, and the phenomenological
    rapidity-scale choice $k=4$. }
    \label{fig:R11_k4}
\end{figure}

A fully consistent treatment would require a dedicated refit of the dipole
initial condition to exclusive data, simultaneously accounting for its
correlation with the evolution prescription. Such an analysis is beyond the
scope of the present work and is left for future investigation. Here, we instead introduce the parameter $k$ as an effective phenomenological prescription to tune the predictions to the data. We use the MV$^\gamma$ initial condition and running-coupling evolution associated with
the fit of Ref.~\cite{Lappi:2013zma}, without the additional collinear
improvement, and find that the choice $k=4$ provides a good
description of both the $Q^2$ dependence and magnitude of the HERA data, as
shown in Fig.~\ref{fig:R11_k4}. The value $k=4$ represents a moderate variation of the rapidity scale and
should be interpreted as an effective phenomenological matching prescription,
rather than as a refit of the dipole amplitude. The fact that such a scale variation substantially improves the description of the
exclusive observable, while the different inclusive-DIS prescriptions display
a common offset, further supports the need for a dedicated refit of the
initial condition to exclusive data. In the remainder of this work, we adopt
this prescription.

\subsection{Helicity-amplitude ratio and spin-density matrix element $r_{00}^{04}$ at HERA and the EIC}

In the remainder of this analysis, we adopt the initial condition and corresponding running-coupling evolution of Ref.~\cite{Lappi:2013zma}, with $x_0=0.01$, supplemented by the phenomenological rapidity-scale choice $k=4$ discussed in the previous subsection. This prescription is used consistently for all the results presented below. Having fixed the small-$x$ input in this way, we now compare the resulting predictions with HERA data and subsequently investigate the corresponding phenomenology for a lead target at the EIC.  \\

\begin{figure}
    \centering
    \includegraphics[width=0.45\linewidth]{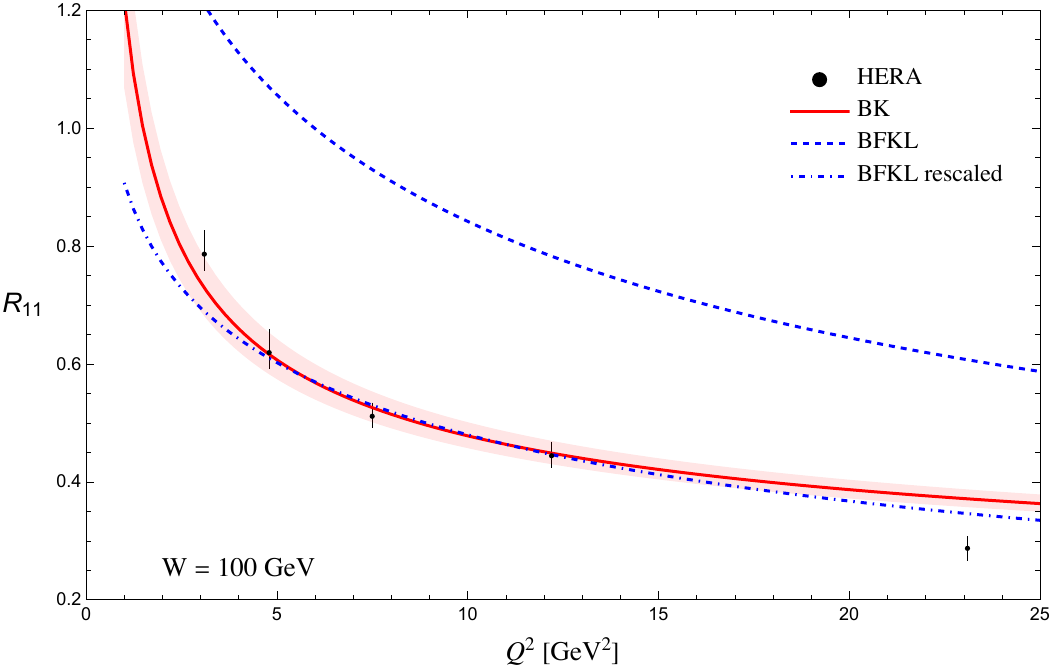} \hspace{0.7 cm}
    \includegraphics[width=0.45\linewidth]{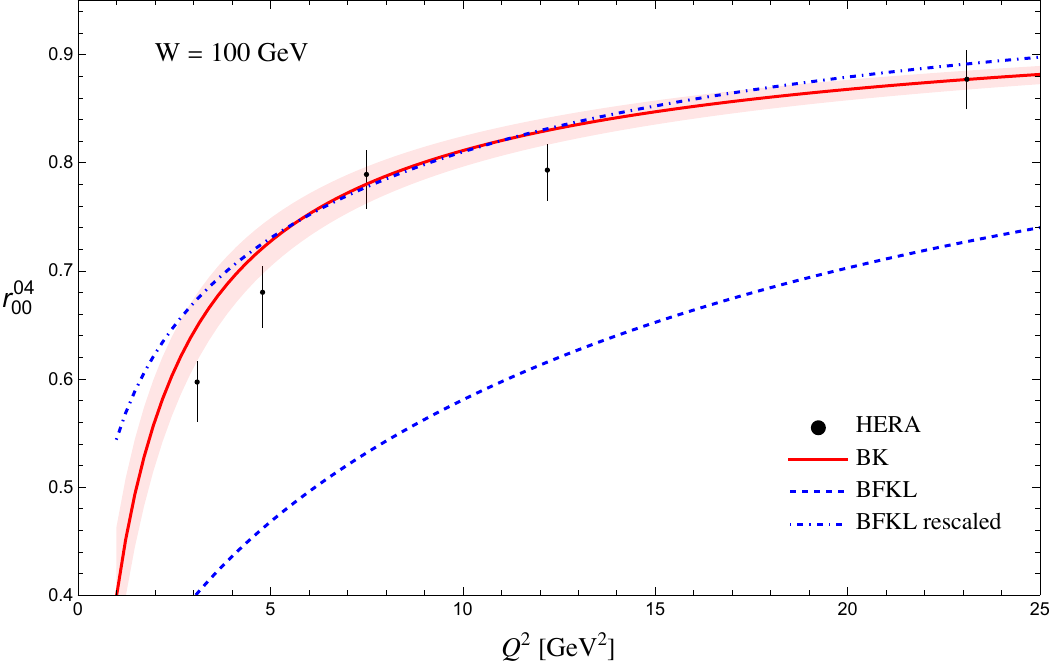}
    \caption{Predictions for the helicity-amplitude ratio $R_{11}$ (left) and for the spin-density matrix element $r_{00}^{04}$ (right), as a function of $Q^2$ at fixed $W=100$ GeV, are compared to the data of H1~\cite{Aaron:2009xp}. The red curves denote predictions obtained by evolving the initial condition in eq.~(\ref{Eq:MV}) through the non-linear BK evolution. The dashed-dot blue line is obtained by evolving the initial condition in eq.~(\ref{Eq:MV}) through the linear BFKL evolution and by fixing its normalization relative to the BK prediction at $Q^2=10~\mathrm{GeV}^2$, where nonlinear-evolution effects are expected to be small. The error bands around the non-linear prediction show only the uncertainty due to the non-perturbative parameters of the DAs.}
    \label{BKvsBFKL}
\end{figure}

In Fig.~\ref{BKvsBFKL}, we compare our results with the data of the H1 experiment~\cite{Aaron:2009xp} at HERA at fixed $W =100$ GeV. We study both linear (BFKL) and non-linear (BK) small-$x$ evolution.   \\

The parameters entering the initial condition were originally fitted to DIS
data using BK evolution~\cite{Lappi:2013zma}. A fully consistent comparison would therefore require
an independent fit of the initial condition using linear BFKL evolution.
In the absence of such a fit, we readjust the BFKL prediction by fixing its
normalization at $Q^2=10~\mathrm{GeV}^2$, where nonlinear-evolution effects are
expected to be already small. This rescaling should be regarded as an
effective, but non-optimal, prescription designed to reduce the normalization
mismatch and facilitate a comparison of the shapes generated by the two
evolution equations. It does not enforce agreement over the entire
moderate-to-high-$Q^2$ region. The most pronounced difference in shape is
found at low $Q^2$, where nonlinear evolution is expected to become relevant, while the residual mismatch at larger $Q^2$ is small and should not be assigned a physical interpretation. \\

As shown in Fig.~\ref{BKvsBFKL}, the BK evolution leads to a modification of both $R_{11}$ and $r_{00}^{04}$ in the low-$Q^2$ region, while the two approaches converge at large $Q^2$, consistently with expectations. This suggests that helicity amplitude ratios and spin-density matrix elements are sensitive to saturation dynamics in exclusive processes. The observables under consideration are ratios of helicity amplitudes and related spin-density matrix elements. Being ratios, they are not a priori guaranteed to exhibit a sensitivity to saturation dynamics, since part of the underlying gluon-density dependence may cancel. For this reason, it is non-trivial to assess whether the non-linear (saturation) effects leave a visible imprint in the low-$Q^2$ region, where they are expected to be most relevant. Quantitatively, the linear and nonlinear predictions begin to separate around 5 GeV$^2$, reaching a 12\% difference at 2 GeV$^2$. A pronounced difference (25\%) is observed only around 1 GeV$^2$, which is similar to the saturation scale expected for a proton at this center-of-mass energy. \\

As noted above, the BFKL result has been rescaled by an overall factor in order to match the BK prediction at a sufficiently large value of $Q^2$, where saturation effects are expected to be negligible. We have chosen $Q^2 = 10\,\mathrm{GeV}^2$, which is roughly an order of magnitude larger than the typical saturation scale for a proton in HERA kinematics. However, as is evident from Figs.~\ref{BKvsBFKL} and~\ref{Fig:Lead_Targ}, the two predictions are slightly misaligned, such that a simple normalization at a fixed $Q^2$ cannot ensure agreement over the entire high-$Q^2$ region. That said, the comparison between BK and BFKL is only non-trivial at low $Q^2$, where this high-$Q^2$ mismatch leads at most to an effect of a few percent,
 which is negligible at the level of accuracy considered here\footnote{In particular, in the absence of full NLO corrections and without the fit mentioned in section \ref{sec:theoretical_uncertainties}.}. In any case, the small discrepancies observed at high $Q^2$ (or moderate $W^2$) between BFKL and BK in our plots should not be interpreted as physical, but rather as an artifact of this normalization procedure. \\

The uncertainty associated with the nonperturbative parameters of the meson
DAs is displayed in Fig.~\ref{BKvsBFKL} only, in order to illustrate that
its size is modest compared with the uncertainty associated with the dipole
initial condition and evolution prescription. \\

\begin{figure}
    \centering
    \includegraphics[width=0.45\linewidth]{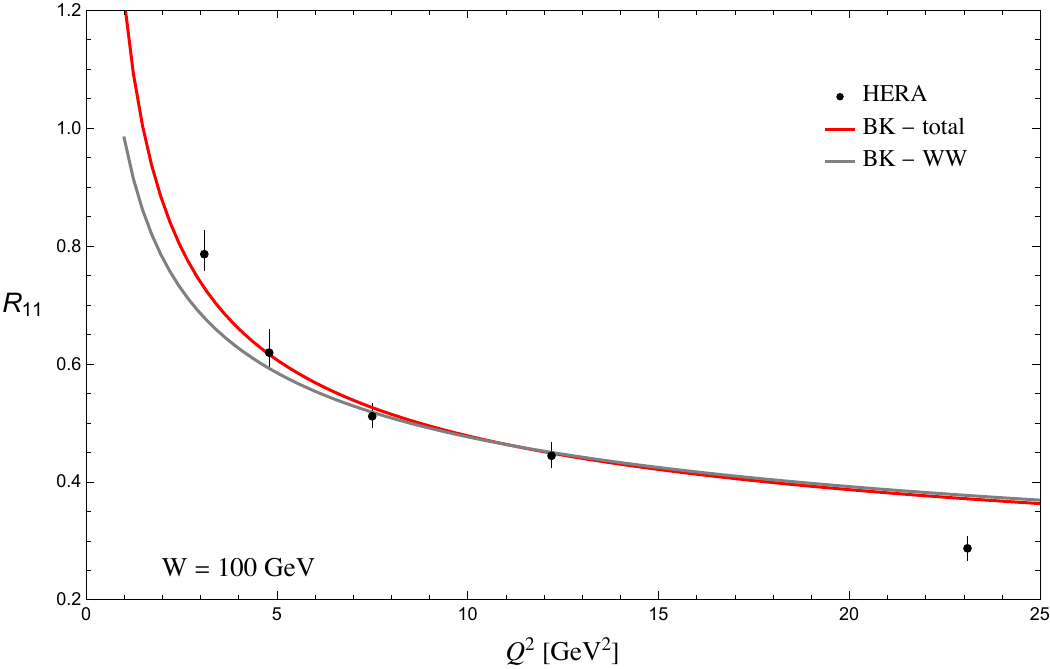} \hspace{0.7 cm}
    \includegraphics[width=0.45\linewidth]{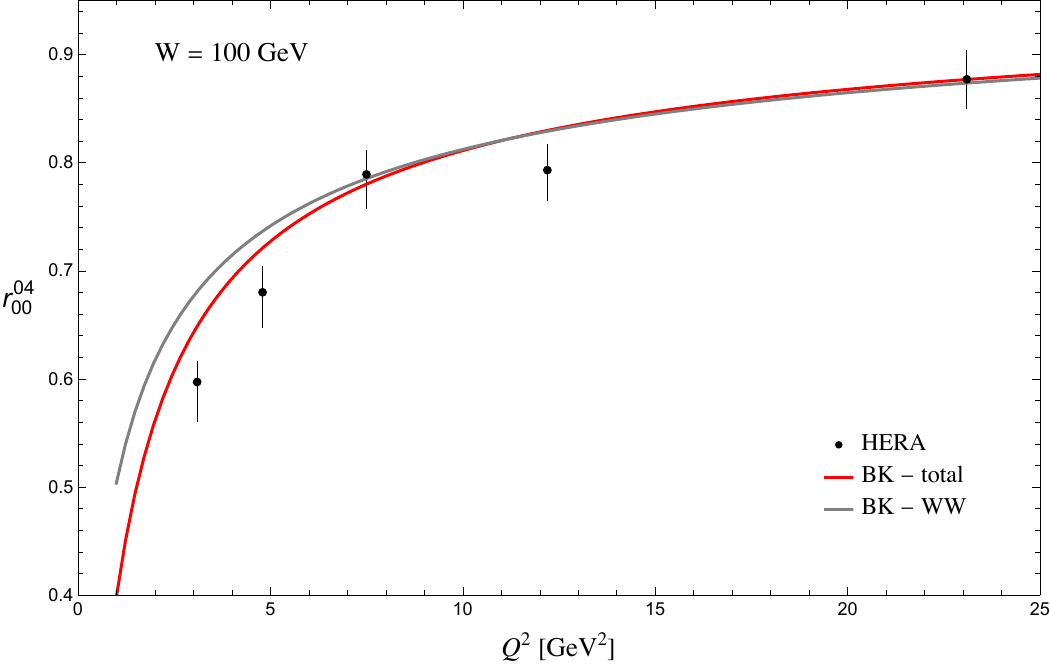}
    \caption{Predictions for the helicity amplitudes ratio $R_{11}$ (left) and for the spin-density matrix element $r_{00}^{04}$ (right), as a function of $Q^2$ at fixed $W=100$ GeV, are compared to the data of H1~\cite{Aaron:2009xp}. Both curves are obtained by evolving the initial condition in eq.~(\ref{Eq:MV}) through the non-linear BK evolution.}
    \label{2BodyVs3body}
\end{figure}

As is well known, at the energies accessible at present colliders the saturation scale lies in the perturbative domain~\cite{AbdulKhalek:2021gbh}. This opens the possibility of investigating the nonlinear regime of QCD using the perturbative tools of the semi-classical small-$x$ effective theory. However, with the exception of fully inclusive processes—most notably DIS—phenomenologically relevant observables sensitive to gluon saturation are typically described within the so-called hybrid factorization framework. In this approach, non-perturbative inputs encoding the partonic structure of hadrons or the hadronization of the final state ({\it e.g.}, fragmentation functions or distribution amplitudes) are introduced. In doing so, genuine higher-twist contributions— such as the three-body component of the DAs considered in this work —are systematically neglected. At most, kinematic twist corrections are included, retaining the transverse-momentum dependence of the lowest Fock states. \\

At present collider energies, the saturation scale remains relatively small. Since sensitivity to non-linear dynamics requires probing virtualities $Q \lesssim Q_s$, phenomenological studies of saturation are necessarily restricted to the low-$Q^2$ region. In this same kinematic regime, power-suppressed contributions are not parametrically small and genuine higher-twist effects, associated with higher Fock components, cannot be neglected \textit{a priori}. Since our calculation consistently incorporates both genuine and kinematic twist-3 terms, it allows us to assess their relative importance at low $Q^2$. Figure~\ref{2BodyVs3body} illustrates the impact of removing the genuine twist-3 contribution: the effect on the helicity ratio and on the spin-density matrix element $r_{04}^{00}$ is typically of the order of that of saturation. It becomes visible below 10 GeV$^2$ and reaches 12\% at 2 GeV$^2$, while it is the same size as non-linear evolution effects at 1 GeV$^2$. Although observables whose leading contribution arises at twist-2 are expected to be less sensitive to genuine higher-twist effects, these contributions are nevertheless not negligible in view of precise measurements of gluon saturation. For instance, from Fig.~\ref{2BodyVs3body}, one can estimate that the impact on the total cross section ($ \sigma_{\rm tot}\sim |A_{LL}|^2+|A_{TT}|^2 $) ranges from approximately 16\% down to about a few \% in the interval $5~\mathrm{GeV}^2 \ge Q^2 \ge 1~\mathrm{GeV}^2$. \\

We now present predictions for DVMP at the EIC in electron–lead collisions. Figure~\ref{Fig:Lead_Targ} shows $R_{11}$ and $r_{00}^{04}$ for a lead target as functions of $Q^2$ at fixed $ W = 140$ GeV. As expected, saturation effects are more pronounced for the lead target, although the increase is relatively modest. At $Q^2 = 1$ GeV$^2$, the magnitude of the saturation effects is approximately 35\% in Fig.~\ref{Fig:Lead_Targ}, whereas in Fig.~\ref{BKvsBFKL} it was about 25\%. Figure~\ref{Fig:Lead_Targ_W} shows the trend in energy at fixed values of $Q^2$. The flat behavior of the ratios is consistent with what was previously observed at HERA (see Fig. 39 in Ref.~\cite{Aaron:2009xp}). As shown in both plots, the linear and nonlinear predictions differ at small values of $Q^2$. However, this difference gradually diminishes and becomes negligible as $Q^2$ increases. The mild increase in the separation between the BK and BFKL predictions towards the lower end of the displayed $W$ range should not be assigned a physical interpretation. It is most likely a residual artifact of the phenomenological rescaling of the BFKL result, whose normalization is fixed relative to BK at a single reference value of $Q^2$. This difference remains nearly constant across the explored energy range. Although the ratios are practically constant over the considered energy range, the separation between the BK and BFKL predictions reflects the different $r$-dependence generated by the two evolution equations. At the initial rapidity, both evolutions start from the same dipole profile. Once evolved to $\eta>\eta_0$, however, their profiles become different, especially in the large-dipole region, where the nonlinear term in the BK equation limits the growth of the dipole amplitude, whereas linear BFKL evolution does not. This region is probed at low $Q^2$, explaining why the difference between the two predictions is most pronounced at small photon virtualities and gradually decreases as $Q^2$ increases.

\begin{figure}
    \centering
    \includegraphics[width=0.45\linewidth]{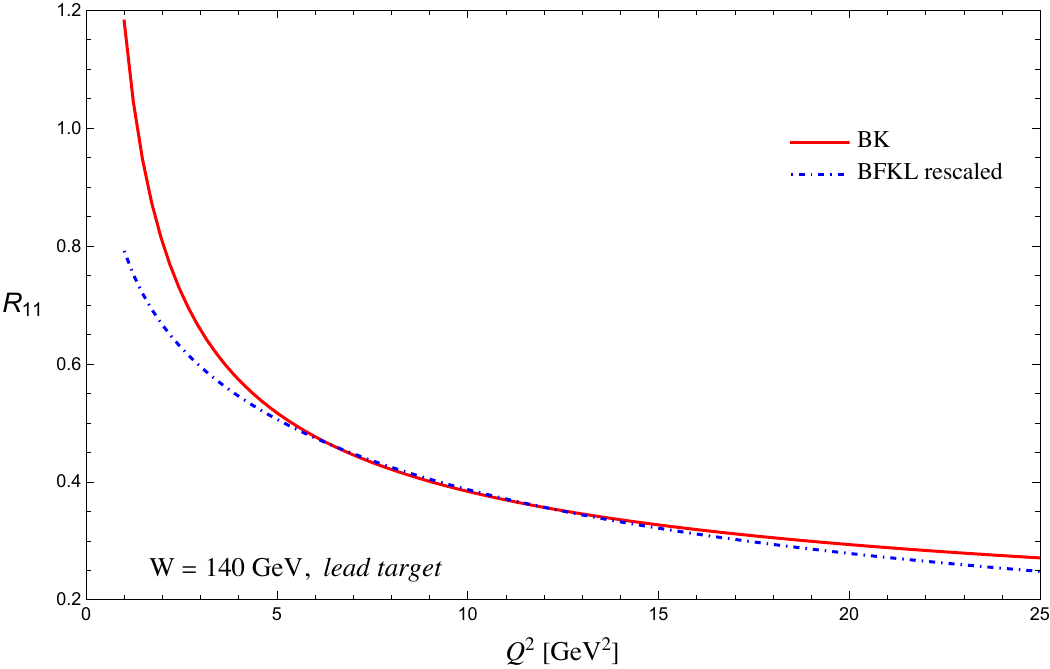} \hspace{0.7 cm}
    \includegraphics[width=0.45\linewidth]{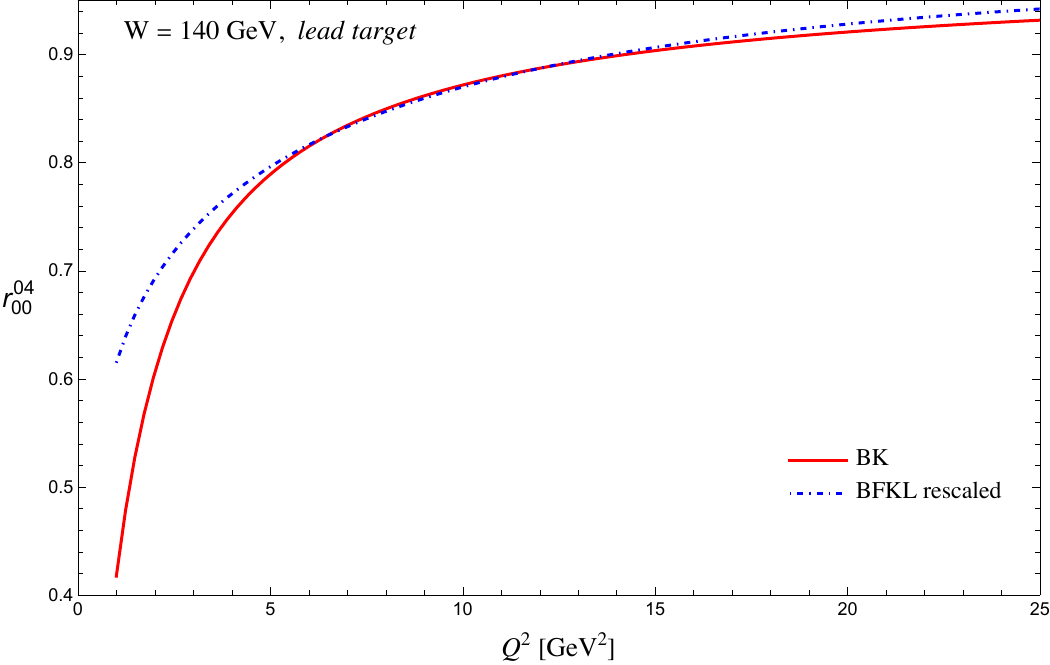}
    \caption{Predictions for the helicity amplitudes ratio $R_{11}$ (left) and for the spin-density matrix element $r_{00}^{04}$ (right), as a function of $Q^2$ at fixed $W=140$ GeV. The red curves denote the predictions obtained by evolving the initial condition in eq.~(\ref{Eq:MV}) through the non-linear BK evolution for a lead target. The dashed-dot blue line is obtained by evolving the initial condition in eq.~(\ref{Eq:MV}) through the linear BFKL evolution and by fixing its normalization relative to the BK prediction at $Q^2=10~\mathrm{GeV}^2$, where nonlinear-evolution effects are expected to be small.}
    \label{Fig:Lead_Targ}
\end{figure}

\begin{figure}
    \centering
    \includegraphics[width=0.45\linewidth]{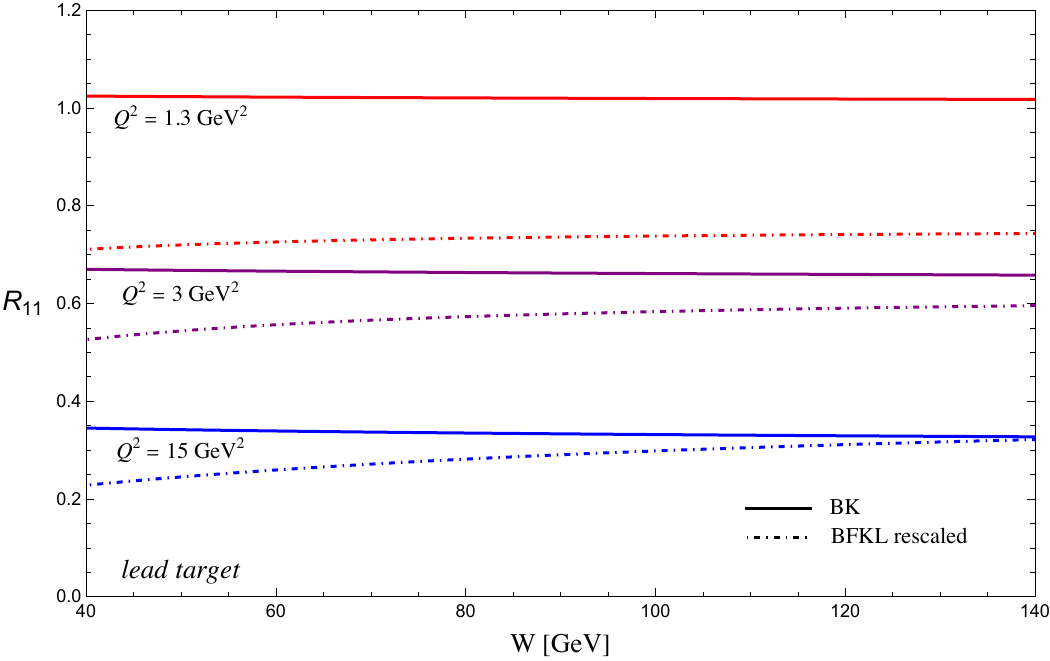} \hspace{0.7 cm}
    \includegraphics[width=0.45\linewidth]{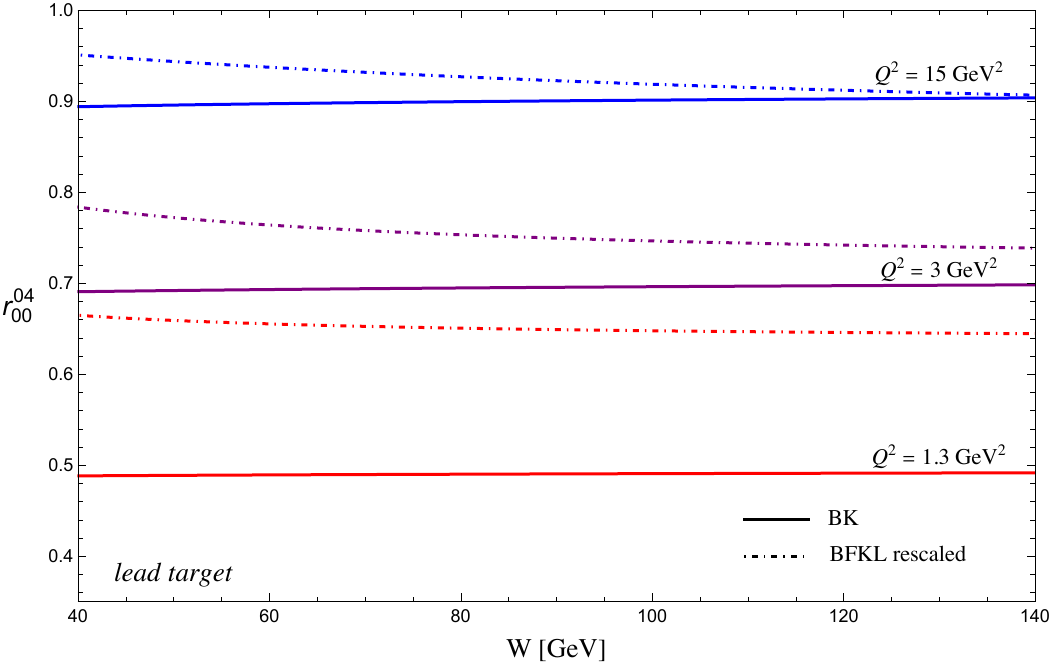}
    \caption{Predictions for the helicity-amplitude ratio $R_{11}$ (left) and the spin-density matrix element $r_{00}^{04}$ (right), as functions of $W$ for a lead target at $Q^2=1.3\,\mathrm{GeV}^2$ (red), $Q^2=3\,\mathrm{GeV}^2$ (purple), and $Q^2=15\,\mathrm{GeV}^2$ (blue). Solid lines correspond to predictions obtained by evolving the initial condition in Eq.~(2.18) through nonlinear BK evolution. Dash-dotted lines correspond to linear BFKL evolution, with the normalization fixed relative to the BK prediction at $Q^2=10\,\mathrm{GeV}^2$, where nonlinear evolution effects are expected to be small.}
    \label{Fig:Lead_Targ_W}
\end{figure}

\section{Conclusions and Outlook}
\label{sec:conclusions}

In this work, we have extended our previous analysis of Deeply Virtual Meson Production by performing a phenomenological study within the Color Glass Condensate framework. We have investigated two helicity-sensitive observables, the transverse-to-longitudinal amplitude ratio
$R_{11}=\mathcal{A}^{11}/\mathcal{A}^{00}$ and the spin-density matrix element $r_{00}^{04}$, including small-$x$ evolution through numerical solutions of the running-coupling BK equation and its linear BFKL limit, with MV- and MV$^\gamma$-type initial conditions. \\

A central aspect of the present analysis is the assessment of the theoretical uncertainty associated with the correlated choice of the dipole initial condition and small-$x$ evolution prescription. We have compared several representative combinations of initial conditions extracted from inclusive DIS fits, running-coupling prescriptions, and evolution equations with and without collinear improvement. Despite a sizeable quantitative spread, these prescriptions produce a similar shape in $Q^2$, broadly reproducing the trend of the HERA data, while systematically overestimating the magnitude of $R_{11}$. This discrepancy cannot therefore be removed solely by switching the collinear improvement on or off or by changing among the considered inclusive-DIS parametrizations. It instead indicates that a dedicated refit of the dipole initial condition to exclusive observables is ultimately required. \\

Since such a refit is beyond the scope of the present work, we have introduced the parameter $k$ as an effective phenomenological prescription to tune the predictions to the data. Using the MV$^\gamma$ initial condition of Ref.~\cite{Lappi:2013zma}, with $x_0=0.01$, running-coupling evolution without additional collinear improvement, and the choice $k=4$, we obtain a good description of both the $Q^2$ dependence and the magnitude of the HERA data. This prescription should not be interpreted as a refit of the dipole amplitude. Rather, its phenomenological success further supports the need for a dedicated determination of the initial condition from exclusive data. \\

Within this common prescription, the predictions obtained with non-linear BK evolution and its linear BFKL limit remain close at moderate and large $Q^2$, after matching their normalizations at $Q^2=10~\mathrm{GeV}^2$, while a more pronounced difference emerges at low $Q^2$. Although the considered quantities are ratios of helicity amplitudes, and therefore exhibit a partial cancellation of the target dependence, they retain a non-trivial sensitivity to the different transverse dipole profiles generated by linear and non-linear evolution. The observed separation should nevertheless be regarded as an indication of their potential sensitivity to non-linear dynamics, rather than as a definitive signal of gluon saturation, given the phenomenological matching and the present theoretical uncertainties. \\

Furthermore, thanks to the consistent inclusion of both kinematic and genuine twist-3 contributions, we have quantified the role of higher-Fock-state components of the light-vector-meson distribution amplitudes. We find that genuine higher-twist effects are sizeable in the low-$Q^2$ region and can be comparable to those associated with non-linear evolution. Their systematic inclusion is therefore essential for a controlled interpretation of helicity observables as probes of small-$x$ dynamics. \\

We have also provided exploratory predictions for electron--lead collisions at the future Electron--Ion Collider, shown in Figs.~\ref{Fig:Lead_Targ} and~\ref{Fig:Lead_Targ_W}. Within the adopted prescription, the separation between linear and non-linear evolution is moderately enhanced for a heavy nuclear target at low $Q^2$, while the two predictions become close at larger virtualities. These results demonstrate the potential discriminating power of the considered observables at the EIC, although their quantitative interpretation will ultimately require a dedicated determination of the initial condition. \\

From the theoretical side, a natural next step is to extend the present analysis to full next-to-leading logarithmic accuracy by including the currently unknown NLO corrections to the transverse impact factors\footnote{The NLO corrections are presently known only for the longitudinal-to-longitudinal transition~\cite{Ivanov:2004pp,Boussarie:2016bkq,Mantysaari:2022bsp}, which is not sufficient for an NLO description of the observables considered here.}. These corrections are particularly important at $Q^2$ of a few GeV$^2$, where non-logarithmically enhanced terms of order $\alpha_s/\pi$ may be sizeable and are required before drawing definitive conclusions about the low-$Q^2$ behavior. \\

The main phenomenological development motivated by the present study is a dedicated fit of the dipole initial condition to exclusive data, consistently accounting for its correlation with the evolution prescription. This analysis could be extended to the complete set of spin-density matrix elements in $\rho$-meson production derived in Refs.~\cite{Boussarie:2024bdo,Boussarie:2024pax}, providing more stringent tests of small-$x$ dynamics and of the interplay between non-linear evolution and higher-twist effects at the EIC. 

\section*{Acknowledgments}
We thank Guillaume Beuf, Piotr Korcyl, and Leszek Motyka for useful discussions. We thank the anonymous referee for comments that helped us improve the quality of this paper. The work of L.D.R. and A.P. is supported by the INFN/QFT@COLLIDERS project (Italy) and by the European Union – Next Generation EU
through the research grant number P2022Z4P4B “SOPHYA - Sustainable Optimised PHYsics
Algorithms: fundamental physics to build an advanced society” under the program PRIN 2022
PNRR of the Italian Ministero dell’Universit\`{a} e Ricerca. The work of L.D.R. has also been supported by the research grant number 20227S3M3B “Bubble Dynamics in Cosmological Phase Transitions” under the program PRIN 2022 of the Italian Ministero dell’Universit\`{a} e Ricerca. The work of M.F. is supported by the ULAM fellowship program of NAWA No. BNI/ULM/2024/1/00065 ``Color glass condensate effective theory beyond the eikonal approximation''. L. S. was supported by the Grant
No. 2024/53/B/ST2/00968 of the National Science Center in Poland.

A.P., L.D.R., L.S. and M.F. gratefully acknowledge the warm hospitality and financial support of IJCLab, where part of this work was carried out. This work was supported by the "P2I - Graduate School of Physics", in the framework ``Investissements d’Avenir'' (ANR-11-IDEX-0003-01) managed by the Agence Nationale de la Recherche (ANR), France. This work was also supported in part by the European Union’s Horizon 2020 research and innovation program under Grant Agreement No. 824093 (Strong2020). This project has also received funding from the French Agence Nationale de la Recherche (ANR) via the grant ANR-20-CE31-0015 (``PrecisOnium'')  and was also partly supported by the French CNRS via the COPIN-IN2P3 bilateral agreement and via the IN2P3 project “QCDFactorisation@NLO”.

\appendix

\section{Evolutions of DAs as a function of $\mu$}
\label{App:RenGroup}

The parameters entering the DAs at $\mu=1$ GeV and their evolution equations  are given in Ref.~\cite{Ball:1998sk}. In Table~\ref{table:DAs} we recall their values for the $\rho$ meson at $\mu_0=1$ GeV.
\begin{table}[htbp]
	\centering
 \begin{tabular}{|l|c|r|}
\hline
 $\alpha_s$ & 0.52\\ \hline
& \\
$\omega^{A}_{\{1,0\}}$ & $-2.1$\\ \hline
&\\
$\omega^{V}_{[0,1]}$ & 28/3\\ \hline
&\\
$a_{2,\rho}$ & 0.18 $\pm$ 0.10\\ \hline
&\\
$m_\rho \,f^{A}_{3\rho}$ & 0.5-0.6 $10^{-2} \,{\rm GeV}^2$\\ \hline
&\\
$m_\rho \, f^{V}_{3\rho}$ & $0.2 \:10^{-2} \,{\rm GeV}^2$\\ \hline
&\\
$f^{A}_{3\rho}/f_{\rho}$ & 0.032\\ \hline
&\\
$f^{V}_{3\rho}/f_{\rho}$ & 0.013\\ \hline
\end{tabular}
\caption{Coupling constants and Gegenbauer coefficients entering the $\rho-$meson DAs, at the scale $\mu_0=1$ GeV. Please note that in Ref.~\cite{Ball:1998sk} the normalization is such that $f^{V,A}_{3\rho\, \mbox{\cite{Ball:1998sk}}}=m_\rho \, f^{V,A}_{3\rho\, [{\rm here}]}$ and $\omega^{V}_{[0,1]}$ and $f_{\rho}$ are fixed constant.}
\label{table:DAs}
\end{table}

\noindent For $a_2$, the evolution equation is
\begin{equation}
a_2(\mu^2)= a_{2}(\mu_0^2) \, L(\mu^2)^{\gamma_2/b_0}
\;, \end{equation}
with
\begin{equation}
\label{defL}
 L(\mu^2) = \frac{\alpha_s(\mu^2)}{\alpha_{s}(\mu_0^2)} = \frac{1}{1+\frac{b_0}{4\pi} \alpha_{s}(\mu_0^2) \ln(\mu^2/\mu_0^2)} \;,
\end{equation}
where $b_0=(11 N_c-2 N_f)/3\,,$
$\gamma_n=4 C_F\left (\psi(n+2)+\gamma_E-\frac{3}{4}-\frac{1}{2 (n+1)(n+2)}\right)$
and
$\psi(n) = -\gamma_E + \sum^{n-1}_{k=1} 1/k\,.$
For the $f^{A}_{3\rho}$ coupling constant, the evolution is given by
\begin{equation}
 f^{A}_{3\rho}(\mu^2)=f^{A}_{3\rho}(\mu_0^2) \, L(\mu^2)^{\Gamma^{-}_2/b_0} \;,
 \end{equation}
with $\Gamma_2^- = -\frac{C_F}{3}+3 \, C_g$ ($C_g = N_c$).
The couplings
$f^{V}_{3\rho} (\mu^2)$ and $\omega^{A}_{ \{ 0,1\} }(\mu^2) f^{A}_{3\rho}(\mu^2)$ enter a matrix evolution equation \cite{Ball:1998sk}. Defining
\beqa
\label{defV}
 V(\mu^2)=\left(
 \begin{array}{clrr}
       \omega^{V}_{[0,1]} f^{V}_{3\rho}(\mu^2) -  \omega^{A}_{ \{ 0,1\} }(\mu^2) f^{A}_{3\rho}(\mu^2)\\
\\
       \omega^{V}_{[0,1]} f^{V}_{3\rho}(\mu^2) + \omega^{A}_{ \{ 0,1\} }(\mu^2) f^{A}_{3\rho}(\mu^2)
\end{array}
\right) \;,
\eqa
it reads
\begin{equation}
\label{V_ev}
V(\mu^2)=L(\mu^2)^{\Gamma^{+}_3/b_0} V(1) \;,
\end{equation}
with $\Gamma^{+}_3$ given by
\beq
\label{defGamma3}
 \Gamma^{+}_3=\left(
 \begin{array}{clrr}
       \frac{8}{3}C_F+\frac{7}{3}C_g \ \ & \ \ \frac{2}{3}C_F-\frac{2}{3}C_g \\
\\
       \frac{5}{3}C_F-\frac{4}{3}C_g \ \ & \ \ \frac{1}{6}C_F+4 C_g
\end{array}
\right)\,.
\eq
Hence, by diagonalizing the system, we can determine the scale dependence of $f^{V}_{3\rho}(\mu^2)$ and $\omega^{A}_{\{0,1\}}(\mu^2)$, thereby obtaining the running of all four coupling constants.

\section{Explicit parametrization of the set of DAs}
\label{App:ExpParam}
The independent DAs are given in eqs.~(\ref{Eq:IndipDALon}),~(\ref{Eq:IndipDAVect3}), and~(\ref{Eq:IndipDAAxial3}), we report them here for convenience of the reader\footnote{We also employ the definitions in eq.~(\ref{Eq:Zeta}).}
\begin{equation*}
    \varphi_{\parallel} (z) = 6 z (1-z) \left( 1 + \frac{3}{2} a_2 \left( \mu^2 \right) (5 ( 2z - 1 )^2 -1) \right) \; ,
\end{equation*}
\begin{equation*}
    V(x_1, x_2) = 5040 \; x_1 x_2 (x_1 - x_2) (1-x_1 - x_2)^2  \; , 
\end{equation*}
\begin{equation*}
     A (x_1, x_2) = 360 x_1 x_2  \left( 1 + \frac{ \omega^A_{\{ 1,0\} } ( \mu^2 ) }{2} (4 - 7 (x_1 + x_2) ) \right) (1-x_1 - x_2)^2 \; .
\end{equation*}
The result in Ref.~\cite{Boussarie:2024bdo} is expressed in terms of DAs that can be explicitly constructed from these three; this is the scope of the present appendix.

\paragraph{Vector DAs.} For the vector part, in Ref.~\cite{Boussarie:2024bdo}, the 2-body twist-3 corrections are expressed in terms of the combination of DAs: $\Tilde{h} (z)  - h(z)$, which naturally emerge in the context of the Covariant Collinear Factorization (CCF)~\cite{Ball:1998sk}. To build an explicit parametrization, we can exploit the relation to the set of DAs derived in the context of Light-Cone Collinear Factorization (LCCF)~\cite{Anikin:2009bf}\footnote{See eq.~(B.29) in Ref.~\cite{Boussarie:2024bdo}.}
\begin{equation}
    \Tilde{h} (z)  - h(z)  = \varphi_1^T (z) \equiv \varphi_1^T (z) \big |_{ \rm W. W.} + \varphi_1^T (z) \big |_{ \rm G. T.}  ,
\end{equation}
where (see eq.~(88) in Ref.~\cite{Anikin:2009bf})
\begin{gather}
    \varphi_1^T (z) \big |_{ \rm W. W.} = \frac{1}{2} \left[ - \bar{z} \int_0^z d u \frac{ \varphi_{\parallel} ( u ) }{\bar{u}} + z \int_z^1 d u \frac{ \varphi_{\parallel} ( u  ) } {u}\right] \nonumber \\
    = \frac{3}{2} z \left(2 z^2-3 z+1\right)
   \bigg( 1 + a_2 \left( \mu^2 \right) ( 1 + 15 z (z-1) ) \bigg) \; ,
\end{gather}
and (see eq.~(97) in Ref.~\cite{Anikin:2009bf} and eq.~(\ref{Eq:BVDARel}) here)
\begin{gather}
    \varphi_1^T (z) \big |_{ \rm G. T.} = \int_0^z dy \; \varphi_3^{\rm gen} (y)  \nonumber \\ +   \zeta_{3M}^V \int_{0}^z d x_1 \int_0^{1-z} d x_2 \frac{V (x_1, x_2)}{(1-x_1-x_2)^2} = \int_0^z dy \; \varphi_3^{\rm gen} (y)  +  840 \zeta_{3M}^V z^2 (1-z)^2 (2 z -1)  \; ,
\end{gather}
where (see eq.~(96) in Ref.~\cite{Anikin:2009bf})
\begin{gather} 
\varphi_3^{\text {gen }}(y) = -\frac{1}{2} \int_y^1 \frac{d u}{u}\left[\int_0^u d y_2 \frac{d}{d u}\left(\zeta_{3M}^V B\left(y_2, u\right) -\zeta_{3M}^A D\left(y_2, u\right) \right)\right.  \nonumber \\  \left.-\int_u^1 \frac{d y_2}{y_2-u}\left(\zeta_{3M}^V B\left(u, y_2\right) -\zeta_{3M}^A D\left(u, y_2\right) \right)-\int_0^u \frac{d y_2}{y_2-u}\left(\zeta_{3M}^V B\left(y_2, u\right) -\zeta_{3M}^A D\left(y_2, u\right) \right)\right]  \nonumber \\ -\frac{1}{2} \int_0^{y} \frac{d u}{\bar{u}}\left[\int_u^1 d y_2 \frac{d}{d u}\left(\zeta_{3M}^V B\left(u, y_2\right)+\zeta_{3M}^A D\left(u, y_2\right) \right)\right.  \nonumber \\  \left.-\int_u^1 \frac{d y_2}{y_2-u}\left(\zeta_{3M}^V B\left(u, y_2\right)+\zeta_{3M}^A D\left(u, y_2\right)\right)-\int_0^u \frac{d y_2}{y_2-u}\left(\zeta_{3M}^V B\left(y_2, u\right) +\zeta_{3M}^A D\left(y_2, u\right) \right) \right] \; ,
\end{gather}
with
\begin{gather}
    B\left(y_1, y_2 \right) = - \frac{V(y_1,1-y_2)}{y_2-y_1} \; , \hspace{1 cm}  D\left(y_1, y_2 \right) = - \frac{A(y_1,1-y_2)}{y_2-y_1} \; .
    \label{Eq:BVDARel}
\end{gather}
Performing the 2-dimensional integrations, we obtain
\begin{gather}
\varphi_3^{\text {gen }}(y) = -\frac{5}{8} \left(3 \omega^A_{\{ 1,0\} } ( \mu^2 ) \left(70 y^4-140 y^3+90 y^2-20 y+1\right) \zeta_{3M}^A -16 \left(6 y^2-6 y+1\right) \zeta_{3M}^A \right. \nonumber \\ \left. -84 \left(70 y^4-140 y^3+90 y^2-20 y+1\right) \zeta_{3M}^V \right) .
\end{gather}
Then, for $\varphi_1^T (z) \big |_{ \rm G. T.}$, we have the parametrization 
\begin{gather}
\varphi_1^T (z) \big |_{ \rm G. T.} = -\frac{5}{8} z \left(2 z^2-3 z+1\right) \bigg[ \zeta_{3M}^A \left(3 \omega^A_{\{ 1,0\} } ( \mu^2 ) \left(7 z^2-7 z+1\right)-16\right) \nonumber \\ -84 \left(23 z^2-23 z+1\right) \zeta_{3M}^V \bigg] \; .
\end{gather}

\paragraph{Axial DAs.} For the axial part, in Ref.~\cite{Boussarie:2024bdo}, the 2-body twist-3 corrections are expressed in terms of the combination of DAs: $ (- g_{\perp}^{(a)}(z) + \Tilde{g}_{\perp}^{(a)} (z) )/4 $. Again, we can exploit the relation to the set of DAs derived in the context of Light-Cone Collinear Factorization (LCCF)~\cite{Anikin:2009bf}\footnote{See eq.~(B.33) in Ref.~\cite{Boussarie:2024bdo}.}
\begin{equation}
        - \frac{ g_{\perp}^{(a)} (z) }{ 4 }  + \frac{\Tilde{g}_{\perp}^{(a)} (z) }{ 4 } =  \varphi_A^T (z) \equiv \varphi_A^T (z) \big |_{ \rm W. W.} + \varphi_A^T (z) \big |_{ \rm G. T.} \; ,
\end{equation}
where (see eq.~(88) in Ref.~\cite{Anikin:2009bf})
\begin{gather}
    \varphi_A^T (z) \big |_{ \rm W. W.} = - \frac{1}{2} \left[ \bar{z} \int_0^z d u \frac{ \varphi_{\parallel} ( u ; \mu^2 ) }{\bar{u}} + z \int_z^1 d u \frac{ \varphi_{\parallel} ( u ; \mu^2 ) }{u}\right] \nonumber \\ = \frac{3}{2} z (z-1) \bigg( 1+ a_2 \left( \mu^2 \right) (1+ 5 (z-1) z) \bigg) \; ,
\end{gather}
and (see eq.~(99) in Ref.~\cite{Anikin:2009bf} and eq.~(\ref{Eq:BVDARel}) here)
\begin{gather}
    \varphi_A^T (z) \big |_{ \rm G. T.} = \int_0^z d y \; \varphi_A^{\rm gen} (y) +  \zeta_{3M}^A \int_{0}^z d x_1 \int_0^{1-z} d x_2 \frac{A (x_1, x_2)}{(1-x_1-x_2)^2} \nonumber \\  = \int_0^z d y \; \varphi_A^{\rm gen} (y ) + 30 \zeta_{3M}^A \bigg( 3 - \omega^A_{\{ 1,0\} } ( \mu^2 ) \bigg) z^2 (1-z)^2 \; ,
\end{gather}
where (see eq.~(98) in Ref.~\cite{Anikin:2009bf})
\begin{equation}
    \varphi_A^{\rm gen} (y) = \varphi_3^{\rm gen} (y) \bigg |_{\zeta_{3M}^V B(x,y) \leftrightarrow \zeta_{3M}^A D(x,y) } \; .
\end{equation}
We can compute $\varphi_A^{\rm gen} (y)$ explicitly and obtain
\begin{gather}
   \varphi_A^{\rm gen} (y) = - \frac{5}{8} \left(20 y^3-30 y^2+12 y-1\right) \bigg[ \bigg( 3 \omega^A_{\{ 1,0\} } ( \mu^2 ) -16 \bigg )  \zeta_{3M}^A -84  \zeta_{3M}^V  \Bigg] \; .
\end{gather}
Then, we get
\begin{gather}
   \varphi_A^T (z) \big |_{ \rm G. T.} = \frac{5}{8} z (1-z) \bigg[ \omega^A_{\{ 1,0\} } ( \mu^2 ) \left(63 z^2-63 z+3\right) \zeta_{3M}^A \nonumber \\ -16 \left(14 z^2-14 z+1\right) \zeta_{3M}^A -84 \left(5 z^2-5 z+1\right) \zeta_{3M}^V \bigg] \; .
\end{gather}

\section{Derivation of the dipole part of the 3-body amplitude}
\label{App:DerivAmp}

In the next subsections, we simplify the expressions obtained in Ref.~\cite{Boussarie:2024bdo}, exploiting the quasi-forward limit. We will consider here only the dipole part of the amplitude. 

\subsection{$ \mathcal{U}_{\boldsymbol{z}_1 \boldsymbol{z}_2} $-part of the amplitude}
\label{sec:Uz1}
We start by considering the {$ \mathcal{U}_{\boldsymbol{z}_1 \boldsymbol{z}_2} $-part of the amplitude:
\begin{gather}
    \mathcal{A}_3^{(a)} =   \left( \prod_{i=1}^3 \int_0^1 d x_i \right) \delta \left( 1 - \sum_i^3 x_i \right) \int d^2 \boldsymbol{z}_1 d^2 \boldsymbol{z}_2 e^{i (\boldsymbol{q}-\boldsymbol{p}_M) \cdot (x_1 \boldsymbol{z}_1 + x_2 \boldsymbol{z}_2)} \frac{ \left\langle P\left(p^{\prime}\right)\left| \mathcal{U}_{\boldsymbol{z}_1 \boldsymbol{z}_2}  \right|P\left(p\right)\right\rangle }{N_c^2}  \nonumber \\
   \times \frac{e_{q} m_M }{8 \pi} c_f \hspace{-0.1 cm} \left( \hspace{-0.1 cm} \varepsilon_{q\rho}-\frac{\varepsilon_{q}^{+}}{q^{+}}q_{\rho} \hspace{-0.1 cm} \right)  \hspace{-0.1 cm} \left( \hspace{-0.1 cm} \varepsilon^{* \mu }_{M} - \frac{p_{M}^{\mu}}{p_M^+} \varepsilon_{M}^{*+} \hspace{-0.1 cm} \right) \hspace{-0.1 cm} \delta \hspace{-0.1 cm} \left( \hspace{-0.1 cm} 1 - \frac{p_M^+}{q^+} \hspace{-0.1 cm} \right) \Psi_3^{(a)} \left( \{ x \},  \boldsymbol{z}_{12} \right)    \; ,
\end{gather}
where we defined the effective impact factor 
\begin{gather}
   \Psi_3^{(a)} \left( \{ x \},  \boldsymbol{z}_{12} \right) = \int d^2 \boldsymbol{z}_3 e^{i x_3 (\boldsymbol{q}-\boldsymbol{p}_M) \boldsymbol{z}_3} \bigg \{ f_{3M}^{V} g_{\sigma \mu} V(x_1, x_2) \left[ \left( 4 g_{\perp}^{ \rho \sigma } \frac{x_1 x_2}{1-x_2} \frac{Q}{Z} K_1 (QZ) \right. \right. \nonumber \\ \left. \left. - 4 i \left[ 2 x_{1}q^{\rho}\left(2\frac{x_{2}}{x_{3}}+1\right) g_{\perp}^{\sigma\nu} - \Xi_1^{\alpha \nu \rho \sigma} i \partial_{z_{1\perp} \alpha} \right] \frac{z_{23 \perp \nu}}{\boldsymbol{z}_{23}^{2}} K_0 (QZ) \right) - \left( 1 \leftrightarrow 2 \right) \right] \nonumber \\ 
    - \epsilon_{- + \sigma \beta} f_{3M}^{A} g^{\beta}_{\perp \mu} A (x_1, x_2) \hspace{-0.1 cm} \left[ \left( \hspace{-0.1 cm} 4 \epsilon^{ \sigma \rho + - } \hspace{-0.05 cm} \frac{x_1 x_2}{1-x_2} \frac{Q}{Z} K_1 (QZ) \hspace{-0.1 cm} \right. \right. \nonumber \\ \left. \left. +  \frac{ 4 i }{ \bar{x}_1 } \bigg[ 2 x_1 \bar{x}_1 q^{\rho} \epsilon^{ \nu \sigma + -} - \Xi_2^{\alpha \nu \rho \sigma} i \partial_{z_{1\perp} \alpha} \bigg] \frac{z_{23 \perp \nu}}{\boldsymbol{z}_{23}^{2}} K_0 (QZ) \hspace{-0.1 cm} \right) \hspace{-0.1 cm} + \hspace{-0.1 cm} \left( 1 \leftrightarrow 2 \right) \hspace{-0.05 cm} \right] \hspace{-0.1 cm} \bigg \} .
    \label{Eq:1234}
\end{gather}
To perform the integration over $\boldsymbol{z}_3$ we need the following integrals:
\begin{gather}
    I_1^{(a)} ( \{ x \}, z_{12} ) = \int d^2 \boldsymbol{z}_3 e^{i x_3 (\boldsymbol{q}-\boldsymbol{p}_M) \boldsymbol{z}_3} \frac{Q}{Z} K_1 (QZ) \; , 
    \label{Eq:I1a}
\end{gather}
\begin{gather}
    I_2^{(a)} ( \{ x \}, z_{12} ) = \int d^2 \boldsymbol{z}_3 e^{i x_3 (\boldsymbol{q}-\boldsymbol{p}_M) \boldsymbol{z}_3} \frac{z_{23 \perp \nu}}{\boldsymbol{z}_{23}^2} K_0 (QZ)  \; , 
    \label{Eq:I2a}
\end{gather}
\begin{gather}
    I_3^{(a)} ( \{ x \}, z_{12} ) = \int d^2 \boldsymbol{z}_3 e^{i x_3 (\boldsymbol{q}-\boldsymbol{p}_M) \boldsymbol{z}_3} \frac{z_{23 \perp \nu}}{\boldsymbol{z}_{23}^2} \frac{\partial}{ \partial z_{1 \perp}^{\alpha}} K_0 (QZ) = \frac{\partial}{ \partial z_{1 \perp}^{\alpha}} I_2^{(a)} ( \{ x \}, z_{12} )  \; .
    \label{Eq:I3a}
\end{gather}
A couple of observations are in order. First, the symmetry under the exchange $(1 \leftrightarrow 2)$ is preserved upon integration. Therefore, the contribution of the exchanged term can be obtained directly from the explicitly written expression in eq.~(\ref{Eq:1234}). Our final goal is to integrate over $\boldsymbol{z}_3$ and then change variables to
\begin{equation}
     \hspace{0.5 cm} \boldsymbol{r} = \boldsymbol{z}_{1} - \boldsymbol{z}_{2} \; , \hspace{1 cm} \boldsymbol{b} = \frac{x_1 \boldsymbol{z}_1 + x_2 \boldsymbol{z}_2}{x_1 + x_2} \; ,
     \label{Eq:Dipole_Impact_12}
\end{equation}
{\it i.e.} the dipole size and impact parameter characterizing the quark-antiquark dipole. When this is done, the dependence on $\boldsymbol{b}$ results in a trivial exponential factor that can be integrated to obtain transverse momentum conservation, $\delta^{(2)} (\boldsymbol{q}-\boldsymbol{p}_M-\boldsymbol{\Delta})$. Since we are interested in near-to-forward kinematics when calculating the above-mentioned integrals, we can take this into account and, after extracting the correct exponential phase, expand around $\boldsymbol{q} - \boldsymbol{p}_M = \boldsymbol{0}$ to simplify the calculation. For instance, we have
\begin{gather}
    I_1^{(a)} ( \{ x \}, z_{12} ) = e^{i \frac{x_3}{x_1 + x_2} (x_1 \boldsymbol{z}_1 + x_2 \boldsymbol{z}_2 ) (\boldsymbol{q} - \boldsymbol{p}_M)} \nonumber \\ \times \frac{2 \pi}{x_3 (x_1 + x_2)} K_0 \left(  \sqrt{ \frac{ x_1 x_2 }{ x_1 + x_2} Q^2 \boldsymbol{z}_{12}^2 } \sqrt{ 1 + \frac{(\boldsymbol{q}-\boldsymbol{p}_M)^2}{x_3 (x_1 + x_2) Q^2} } \right) \; ,
\end{gather}
that can be simplified to
\begin{gather}
    \frac{ I_1^{(a)} ( \{ x \}, z_{12} ) }{ e^{ i \frac{x_3}{x_1 + x_2} (x_1 \boldsymbol{z}_1 + x_2 \boldsymbol{z}_2 ) (\boldsymbol{q} - \boldsymbol{p}_M)} } \bigg |_{ \rm f.} =  \frac{2 \pi}{x_3 (x_1 + x_2)} K_0 \left(  \sqrt{ \frac{ x_1 x_2 }{ x_1 + x_2} Q^2 \boldsymbol{z}_{12}^2 } \right) 
    \equiv \frac{2 \pi}{x_3} F_1 \left( x_1, x_2, \boldsymbol{z}_{12}^2 \right) \; ,
    \label{Eq:I_a1_Fin}
\end{gather}
observing that this integral enters the $T \rightarrow T$ transition. Please note that, for the transverse-momentum integrals, we introduce the notations \( |_{\rm F.} \) (forward) and \( |_{\rm n.F.} \) (next-to-forward) to indicate the level of accuracy at which the integral is evaluated in the \( \boldsymbol{q} - \boldsymbol{p}_M \) expansion around zero. Moreover, n.F. is also employed as a superscript of some functions appearing in the next-to-forward approximation. This notation should not be confused with the notions of flip (f.) and non-flip (n.f.). This observation is particularly useful for obtaining fully integrated forms for $I_2$ and $I_3$, which read
\begin{gather}
    \frac{I_2^{(a)} ( \{ x \}, z_{12} )}{e^{i \frac{x_3}{x_1 + x_2} (x_1 \boldsymbol{z}_1 + x_2 \boldsymbol{z}_2 ) (\boldsymbol{q} - \boldsymbol{p}_M)}} \bigg |_{\rm n. F.} =  \frac{2 \pi}{x_3} \Bigg \{ \frac{ z_{1 2 \perp \nu} }{ x_1 \boldsymbol{z}_{12}^2 Q^2} \Bigg[ - \sqrt{ \frac{x_1 x_2 }{x_1 + x_2} \boldsymbol{z}_{12}^2 Q^2 } \; K_1 \left(  \sqrt{ \frac{x_1 x_2 }{x_1 + x_2} \boldsymbol{z}_{12}^2 Q^2 }  \right)  \nonumber \\ \left. 
    + \sqrt{x_1 \bar{x}_1 Q^2 \boldsymbol{z}_{12}^2 } K_1 \left( \sqrt{x_1 \bar{x}_1 Q^2 \boldsymbol{z}_{12}^2 } \right) \bigg] + \frac{x_3}{Q^2} \left[ \frac{z_{12 \perp \nu} \; i (q-p_M)_{\perp} \cdot z_{12 \perp} }{ \boldsymbol{z}_{12}^2 } \frac{ \sqrt{x_1 \bar{x}_1 Q^2 \boldsymbol{z}_{12}^2 } }{(x_1+x_2)} \right. \right. \nonumber \\ \left. \left. \times  K_1 \left( \sqrt{x_1 \bar{x}_1 Q^2 \boldsymbol{z}_{12}^2 } \right) + \left( i\left( p_{M} - q \right)_{\perp \nu} - 2 z_{12 \perp \nu} \frac{i (q-p_M)_{\perp } \cdot z_{12 \perp} }{ \boldsymbol{z}_{12}^2 } \right) \right. \right. \nonumber \\ \left. \left. \times  \left( \frac{x_2}{(x_1+x_2) x_1 x_3} K_2 \left( \sqrt{ \frac{x_1 x_2 Q^2 \boldsymbol{z}_{12}^2}{x_1+x_2} } \right) - \frac{x_2 + x_3}{ x_1 x_3} K_2 \left( \sqrt{x_1 \bar{x}_1 Q^2 \boldsymbol{z}_{12}^2 } \right) \right)  \right] \right \} \nonumber \\ \equiv \frac{2 \pi}{x_3} \bigg \{ z_{12 \perp \nu} F_2 \left( x_1, x_2, \boldsymbol{z}_{12}^2 \right) + z_{12 \perp \nu} \; i (q-p_M)_{\perp} \cdot z_{12 \perp} F_2^{ \rm n. F.} \left( x_1, x_2, \boldsymbol{z}_{12}^2 \right) \nonumber \\ - ( \boldsymbol{z}_{12}^2 \; i (q-p_M)_{\perp \nu} + 2 z_{12 \perp \nu} \; i (q-p_M)_{\perp} \cdot z_{12 \perp} ) \tilde{F}_2^{ \rm n. F.} \left( x_1, x_2, \boldsymbol{z}_{12}^2 \right) \bigg \} 
    \label{Eq:I_a2_Fin}
\end{gather}
and 
\begin{gather}
    \frac{I_3^{(a)} ( \{ x \}, z_{12} )}{e^{ i \frac{x_3}{x_1 + x_2} (x_1 \boldsymbol{z}_1 + x_2 \boldsymbol{z}_2 ) (\boldsymbol{q} - \boldsymbol{p}_M)}} \bigg |_{\rm F.} =   
     \frac{2 \pi}{x_1 x_3 Q^2 \boldsymbol{z}_{12}^2} \left(g_{\perp \alpha \nu}+\frac{2 z_{12 \perp \nu} z_{12 \perp \alpha}}{\boldsymbol{z}_{12}^2}\right)  \nonumber \\ \times \left[\sqrt{x_1 \bar{x}_1 Q^2 \boldsymbol{z}_{12}^2} K_1\left(\sqrt{x_1 \bar{x}_1 Q^2 \boldsymbol{z}_{12}^2}\right)-\sqrt{\frac{x_1 x_2}{x_1+x_2} Q^2 \boldsymbol{z}_{12}^2} K_1\left(\sqrt{\frac{x_1 x_2}{x_1+x_2} Q^2 \boldsymbol{z}_{12}^2}\right)\right]  \nonumber \\
    +  \frac{2 \pi}{x_1 x_3} \frac{\boldsymbol{z}_{12 \perp \nu} z_{12 \perp \alpha}}{\boldsymbol{z}_{12}^2} \left[x_1 \bar{x}_1 K_0\left(\sqrt{x_1 \bar{x}_1 Q^2 \boldsymbol{z}_{12}^2}\right)-\frac{x_1 x_2}{x_1+x_2} K_0\left(\sqrt{\frac{x_1 x_2}{x_1+x_2} Q^2 \boldsymbol{z}_{12}^2}\right)\right] \nonumber \\
   = \frac{2 \pi}{x_3} \left[ \left(g_{\perp \alpha \nu}+\frac{2 z_{12 \perp \nu} z_{12 \perp \alpha}}{\boldsymbol{z}_{12}^2}\right) F_3 \left( x_1, x_2 , \boldsymbol{z}_{12}^2  \right) + \frac{\boldsymbol{z}_{12 \perp \nu} z_{12 \perp \alpha}}{\boldsymbol{z}_{12}^2} \tilde{F}_3 \left( x_1, x_2 , \boldsymbol{z}_{12}^2  \right) \right] \; .
   \label{Eq:I_a3_Fin}
\end{gather}
We observe that, in the first integral, we kept a next-to-forward accuracy in order to get a non-zero contribution to the $s$-channel non-conserving helicity amplitude. We observe that we can greatly simplify the expression of the longitudinal-photon-to-transverse-meson amplitude by observing that 
\begin{gather*}
    \int_0^{2 \pi} d \varphi \bigg(  2 (\boldsymbol{\varepsilon}_M^{*} \cdot \boldsymbol{r}) (\boldsymbol{r} \cdot \boldsymbol{\Delta})  - (\boldsymbol{\varepsilon}_M^{*} \cdot \boldsymbol{\Delta}) \; \boldsymbol{r}^2 \bigg) = 0 \; ,
\end{gather*}
where $\varphi$ is the angle associated to the vector $\boldsymbol{r}$. This implies that the structure $\tilde{F}_2^{\rm n.f.}$ will not contribute. \\

Performing the integrals and the tensor contraction and making the change of variables in (\ref{Eq:Dipole_Impact_12}), we can finally obtain 
\begin{gather}
  \mathcal{A}_3^{(a) T, L} =  (2 \pi)^4 W^2 \; \delta^{(4)} (q + p_t - p_M -p_{t'} ) \left( \prod_{i=1}^3 \int_0^1 d x_i \right) \delta \left(1- \sum_i^3 x_i \right) \int d^2 \boldsymbol{r} \; \mathcal{U} (\boldsymbol{r}) \nonumber \\ \times \left( - \frac{e_q m_M}{ \pi} Q^2 (\boldsymbol{r} \cdot \boldsymbol{\Delta}) \right) \left[ \frac{\varepsilon_{\gamma}^{+}}{q^+} (\boldsymbol{\varepsilon}_M^{*} \cdot \boldsymbol{r}) \right]  \Bigg \{  (1-c_f)  \phi_{3 L} (x_1, x_2) \nonumber \\ \times  \bigg( \frac{x_1^2}{x_3 (x_1+x_2)} F_2 (x_1, x_2, \boldsymbol{r}^2)  -  \frac{x_1}{x_3}  F_2^{ \rm n. F.} (x_1, x_2, \boldsymbol{r}^2)  \bigg)  + (x_1 \leftrightarrow x_2) \Bigg \} 
\end{gather}
for the $L \rightarrow T$ transition and 
\begin{gather}
  \mathcal{A}_3^{(a) T, T} =  (2 \pi)^4 W^2 \; \delta^{(4)} (q + p_t - p_M -p_{t'} ) \left( \prod_{i=1}^3 \int_0^1 d x_i \right) \delta \left( 1 - \sum_i^3 x_i \right) \int d^2 \boldsymbol{r} \; \mathcal{U} (\boldsymbol{r}) \nonumber \\ \times \left( - \frac{e_q m_M}{8 \pi} \right) (1-c_f) \Bigg \{  \frac{8}{x_3^2} T_{\rm f.} (\boldsymbol{r}) \bigg(2 F_3 (x_1, x_2, \boldsymbol{r} ) + \tilde{F}_3 (x_1, x_2, \boldsymbol{r} ) \bigg) \phi_{3, (+) } (x_1,x_2) - \frac{4}{x_3} \; T_{\rm n. f.} (\boldsymbol{r}) \nonumber \\  \times \left[ \frac{x_1 x_2}{1 - x_2}  F_1 (x_1, x_2, \boldsymbol{r} ) \phi_{3, (-) } (x_1,x_2)  \right. \nonumber \\ \left.  + \frac{\tilde{F}_3 (x_1, x_2, \boldsymbol{r} )}{(1-x_1) x_3} \bigg( x_1 \phi_{3, (+) } (x_1,x_2) - (1-2 x_1) \phi_{3, (-) } (x_1,x_2) \bigg) \right] + (x_1 \leftrightarrow x_2) \Bigg \} 
\end{gather}
for the $T \rightarrow T$ transition, where we introduced the combination of DAs
\begin{gather}
    \phi_{3, L } (x_1,x_2) = f_{3M}^{A} A (x_1, x_2) - \left( 1 +  \frac{2 x_2}{x_3} \right) f_{3M}^{V} V (x_1, x_2) \; ,
\end{gather}
\begin{gather}
    \phi_{3, (+) } (x_1,x_2) = (x_1+x_2) f_{3M}^{A} A (x_1, x_2) + (x_2-x_1) f_{3M}^{V} V (x_1, x_2) \; ,
\end{gather}
\begin{gather}
    \phi_{3, (-) } (x_1,x_2) = (f_{3M}^V V(x_1, x_2) - f_{3M}^A A(x_1, x_2)) \; .
\end{gather}

\subsection{$ \mathcal{U}_{\boldsymbol{z}_1 \boldsymbol{z}_3} $ plus $\mathcal{U}_{\boldsymbol{z}_3 \boldsymbol{z}_2}$ -part of the amplitude} 
\label{sec:Uz2z3}
It is convenient to consider the gluon-quark dipoles simultaneously:
\begin{gather}
    \mathcal{A}_3^{(b)} = -   \left( \prod_{i=1}^3 \int_0^1 d x_i \right) \delta \left( 1 - \sum_i^3 x_i \right) \frac{e_{q} m_M }{8 \pi} c_f \hspace{-0.1 cm} \left( \hspace{-0.1 cm} \varepsilon_{q\rho}-\frac{\varepsilon_{q}^{+}}{q^{+}}q_{\rho} \hspace{-0.1 cm} \right)  \hspace{-0.1 cm} \left( \hspace{-0.1 cm} \varepsilon^{* \mu }_{M} - \frac{p_{M}^{\mu}}{p_M^+} \varepsilon_{M}^{*+} \hspace{-0.1 cm} \right) \hspace{-0.1 cm} \delta \hspace{-0.1 cm} \left( \hspace{-0.1 cm} 1 - \frac{p_M^+}{q^+} \hspace{-0.1 cm} \right) \nonumber \\ \times \Bigg \{ \int d^2 \boldsymbol{z}_1 d^2 \boldsymbol{z}_3 \; e^{i (\boldsymbol{q}-\boldsymbol{p}_M) (x_1 \boldsymbol{z}_1 + x_3 \boldsymbol{z}_3)}  \left\langle P\left(p^{\prime}\right)\left| \mathcal{U}_{\boldsymbol{z}_1 \boldsymbol{z}_3}  \right|P\left(p\right)\right\rangle \int d^2 \boldsymbol{z}_2 \; e^{i x_2 (\boldsymbol{q}-\boldsymbol{p}_M) \boldsymbol{z}_2}    \Psi_3 \left( \{ x \}, \{ \boldsymbol{z} \} \right) \nonumber \\ + \int d^2 \boldsymbol{z}_2 d^2 \boldsymbol{z}_3 \; e^{i (\boldsymbol{q}-\boldsymbol{p}_M) (x_2 \boldsymbol{z}_2 + x_3 \boldsymbol{z}_3)}  \left\langle P\left(p^{\prime}\right)\left| \mathcal{U}_{\boldsymbol{z}_3 \boldsymbol{z}_2}  \right|P\left(p\right)\right\rangle \int d^2 \boldsymbol{z}_1 \; e^{i x_1 (\boldsymbol{q}-\boldsymbol{p}_M) \boldsymbol{z}_1}  \Psi_3 \left( \{ x \}, \{ \boldsymbol{z} \} \right) \Bigg \} \; .
\end{gather}
In the second term (last line), if we make the relabelling $\boldsymbol{z}_2 \leftrightarrow \boldsymbol{z}_1$ and use the property
\begin{equation}
    \Psi_3 \left( x_1, x_2, x_3 , \boldsymbol{z}_2 , \boldsymbol{z}_1 , \boldsymbol{z}_3 \right) = \Psi_3 \left( x_2, x_1, x_3 , \boldsymbol{z}_1 , \boldsymbol{z}_2, \boldsymbol{z}_3 \right) \;,
\end{equation}
we obtain 
\begin{gather}
    \mathcal{A}_3^{(b)} = -   \left( \prod_{i=1}^3 \int_0^1 d x_i \right) \delta \left( 1 - \sum_i^3 x_i \right) \frac{e_{q} m_M }{8 \pi} c_f \hspace{-0.1 cm} \left( \hspace{-0.1 cm} \varepsilon_{q\rho}-\frac{\varepsilon_{q}^{+}}{q^{+}}q_{\rho} \hspace{-0.1 cm} \right)  \hspace{-0.1 cm} \left( \hspace{-0.1 cm} \varepsilon^{* \mu }_{M} - \frac{p_{M}^{\mu}}{p_M^+} \varepsilon_{M}^{*+} \hspace{-0.1 cm} \right) \hspace{-0.1 cm} \delta \hspace{-0.1 cm} \left( \hspace{-0.1 cm} 1 - \frac{p_M^+}{q^+} \hspace{-0.1 cm} \right) \nonumber \\ \times \int d^2 \boldsymbol{z}_1 d^2 \boldsymbol{z}_3 \Bigg \{  \; e^{i (\boldsymbol{q}-\boldsymbol{p}_M) (x_1 \boldsymbol{z}_1 + x_3 \boldsymbol{z}_3)}  \left\langle P\left(p^{\prime}\right)\left| \mathcal{U}_{\boldsymbol{z}_1 \boldsymbol{z}_3}  \right|P\left(p\right)\right\rangle \Psi_3^{(b)} \left( x_1, x_2, x_3 , \boldsymbol{z}_{13} \right) \nonumber \\ + \; e^{i (\boldsymbol{q}-\boldsymbol{p}_M) (x_2 \boldsymbol{z}_1 + x_3 \boldsymbol{z}_3)}  \left\langle P\left(p^{\prime}\right)\left| \mathcal{U}_{\boldsymbol{z}_3 \boldsymbol{z}_1}  \right|P\left(p\right)\right\rangle \Psi_3^{(b)} \left( x_2, x_1, x_3 , \boldsymbol{z}_{13} \right) \Bigg \} \; ,
\end{gather}
where
\begin{gather}
   \Psi_3^{(b)} \left( \{ x \},  \boldsymbol{z}_{13} \right) = \int d^2 \boldsymbol{z}_2 e^{i x_2 (\boldsymbol{q}-\boldsymbol{p}_M) \boldsymbol{z}_2} \bigg \{ f_{3M}^{V} g_{\sigma \mu} V(x_1, x_2) \left[ \left( 4 g_{\perp}^{ \rho \sigma } \frac{x_1 x_2}{1-x_2} \frac{Q}{Z} K_1 (QZ) \right. \right. \nonumber \\ \left. \left. - 4 i \left[ 2 x_{1}q^{\rho}\left(2\frac{x_{2}}{x_{3}}+1\right) g_{\perp}^{\sigma\nu} - \Xi_1^{\alpha \nu \rho \sigma} i \partial_{z_{1\perp} \alpha} \right] \frac{z_{23 \perp \nu}}{\boldsymbol{z}_{23}^{2}} K_0 (QZ) \right) - \left( 1 \leftrightarrow 2 \right) \right] \nonumber \\ 
    - \epsilon_{- + \sigma \beta} f_{3M}^{A} g^{\beta}_{\perp \mu} A (x_1, x_2) \hspace{-0.1 cm} \left[ \left( \hspace{-0.1 cm} 4 \epsilon^{ \sigma \rho + - } \hspace{-0.05 cm} \frac{x_1 x_2}{1-x_2} \frac{Q}{Z} K_1 (QZ) \hspace{-0.1 cm} \right. \right. \nonumber \\ \left. \left. +  \frac{ 4 i }{ \bar{x}_1 } \bigg[ 2 x_1 \bar{x}_1 q^{\rho} \epsilon^{ \nu \sigma + -} - \Xi_2^{\alpha \nu \rho \sigma} i \partial_{z_{1\perp} \alpha} \bigg] \frac{z_{23 \perp \nu}}{\boldsymbol{z}_{23}^{2}} K_0 (QZ) \hspace{-0.1 cm} \right) \hspace{-0.1 cm} + \hspace{-0.1 cm} \left( 1 \leftrightarrow 2 \right) \hspace{-0.05 cm} \right] \hspace{-0.1 cm} \bigg \} .
\end{gather}
Since the dipole matrix element depends on the relative distance vector $\boldsymbol{z}_{13}$, it is clear that the second contribution is obtained from the first through the exchange $(x_1, x_2, \boldsymbol{z}_{13}) \rightarrow (x_2, x_1, -\boldsymbol{z}_{13})$. Thus, we can write  
\begin{gather}
    \mathcal{A}_3^{(b)} = -   \left( \prod_{i=1}^3 \int_0^1 d x_i \right) \delta \left( 1 - \sum_i^3 x_i \right) \frac{e_{q} m_M }{8 \pi} c_f \hspace{-0.1 cm} \left( \hspace{-0.1 cm} \varepsilon_{q\rho}-\frac{\varepsilon_{q}^{+}}{q^{+}}q_{\rho} \hspace{-0.1 cm} \right)  \hspace{-0.1 cm} \left( \hspace{-0.1 cm} \varepsilon^{* \mu }_{M} - \frac{p_{M}^{\mu}}{p_M^+} \varepsilon_{M}^{*+} \hspace{-0.1 cm} \right) \hspace{-0.1 cm} \delta \hspace{-0.1 cm} \left( \hspace{-0.1 cm} 1 - \frac{p_M^+}{q^+} \hspace{-0.1 cm} \right) \nonumber \\ \times \int d^2 \boldsymbol{z}_1 d^2 \boldsymbol{z}_3 \left\langle P\left(p^{\prime}\right)\left| \mathcal{U}_{\boldsymbol{z}_1 \boldsymbol{z}_3}  \right|P\left(p\right)\right\rangle \Bigg \{  \; e^{i (\boldsymbol{q}-\boldsymbol{p}_M) (x_1 \boldsymbol{z}_1 + x_3 \boldsymbol{z}_3)}   \Psi_3^{(b)} \left( x_1, x_2, x_3 , \boldsymbol{z}_{13} \right) \nonumber \\ + ( x_1 \leftrightarrow x_2, \boldsymbol{z}_{13} \rightarrow - \boldsymbol{z}_{13}) \Bigg \} \; .
\end{gather}
Since we integrate over $\boldsymbol{z}_2$, it is clear that the symmetry (1 $\leftrightarrow$ 2) within the function $\Psi_3^{(b)}$ is not preserved. To integrate all terms explicitly we need the following integrals:
\begin{gather}
    I_1^{(b)} ( \{ x \}, z_{13} ) = \int d^2 \boldsymbol{z}_2 e^{i x_2 (\boldsymbol{q}-\boldsymbol{p}_M) \boldsymbol{z}_2} \frac{Q}{Z} K_1 (QZ) \; ,
    \label{Eq:I1b}
\end{gather}
\begin{gather}
    I_2^{(b)} ( \{ x \}, z_{13} ) = \int d^2 \boldsymbol{z}_2 e^{i x_2 (\boldsymbol{q}-\boldsymbol{p}_M) \boldsymbol{z}_2} \frac{z_{23 \perp \nu}}{\boldsymbol{z}_{23}^2} K_0 (QZ) \; ,
    \label{Eq:I2b}
\end{gather}
\begin{gather}
    I_3^{(b)} ( \{ x \}, z_{13} ) = \int d^2 \boldsymbol{z}_2 e^{i x_2 (\boldsymbol{q}-\boldsymbol{p}_M) \boldsymbol{z}_2} \frac{z_{23 \perp \nu}}{\boldsymbol{z}_{23}^2} \frac{\partial}{ \partial z_{1 \perp}^{\alpha}} K_0 (QZ) = \frac{\partial}{ \partial z_{1 \perp}^{\alpha}} I_1^{(b)} ( \{ x \}, z_{13} ) \; ,
    \label{Eq:I3b}
\end{gather}
\begin{gather}
    I_4^{(b)} ( \{ x \}, z_{13} ) = \int d^2 \boldsymbol{z}_2 e^{i x_2 (\boldsymbol{q}-\boldsymbol{p}_M) \boldsymbol{z}_2} K_0 (QZ) \; . 
    \label{Eq:I4b}
\end{gather}
The first three integrals can be obtained from the ones in eqs.~(\ref{Eq:I1a}), (\ref{Eq:I2a}), (\ref{Eq:I3a}), under suitable exchanges of pair of variables $(x_i, \boldsymbol{z}_i)$. The absence of the integral
\begin{gather}
     \int d^2 \boldsymbol{z}_2 e^{i x_2 (\boldsymbol{q}-\boldsymbol{p}_M) \boldsymbol{z}_2} \frac{z_{13 \perp \nu}}{\boldsymbol{z}_{13}^2} \frac{\partial}{ \partial z_{2 \perp}^{\alpha}} K_0 (QZ)  
\end{gather}
is due to the fact that the latter starts at next-to-forward accuracy and enters the $T \rightarrow T$ transition, thus being subleading in this helicity amplitude. The integrals in eqs.~(\ref{Eq:I1b}), (\ref{Eq:I2b}), (\ref{Eq:I3b}) and (\ref{Eq:I4b}) read
\begin{gather}
    \frac{ I_1^{(b)} ( \{ x \}, z_{13} ) }{ e^{ i \frac{x_2}{x_1 + x_3} (x_1 \boldsymbol{z}_1 + x_3 \boldsymbol{z}_3 ) (\boldsymbol{q} - \boldsymbol{p}_M)} } \bigg |_{ \rm F.} =  \frac{2 \pi}{x_2 (x_1 + x_3)} K_0 \left(  \sqrt{ \frac{ x_1 x_3 }{ x_1 + x_3} Q^2 \boldsymbol{z}_{13}^2 } \right) 
    \equiv \frac{2 \pi}{x_2} F_1 \left( x_1, x_3, \boldsymbol{z}_{13}^2 \right)  \; ,
    \label{Eq:I_b1_Fin}
\end{gather}
\begin{gather}
    \frac{I_2^{(b)} ( \{ x \}, z_{13} )}{e^{i \frac{x_2}{x_1 + x_3} (x_1 \boldsymbol{z}_1 + x_3 \boldsymbol{z}_3 ) (\boldsymbol{q} - \boldsymbol{p}_M)}} \bigg |_{\rm n. F.} =  -\frac{2 \pi}{x_2} \Bigg \{ \frac{ z_{1 3 \perp \nu} }{ x_1 \boldsymbol{z}_{13}^2 Q^2} \Bigg[  -  \sqrt{ \frac{x_1 x_3 }{x_1 + x_3} \boldsymbol{z}_{13}^2 Q^2 } \; K_1 \left(  \sqrt{ \frac{x_1 x_3 }{x_1 + x_3} \boldsymbol{z}_{13}^2 Q^2 }  \right)  \nonumber \\ \left. 
     +  \sqrt{x_1 \bar{x}_1 Q^2 \boldsymbol{z}_{13}^2 } K_1 \left( \sqrt{x_1 \bar{x}_1 Q^2 \boldsymbol{z}_{13}^2 } \right) \bigg] + \frac{x_2}{Q^2} \left[ \frac{z_{13 \perp \nu} \; i (q-p_M)_{\perp} \cdot z_{13 \perp} }{ \boldsymbol{z}_{13}^2 } \frac{ \sqrt{x_1 \bar{x}_1 Q^2 \boldsymbol{z}_{13}^2 } }{(x_1+x_3)} \right. \right. \nonumber \\ \left. \left. \times  K_1 \left( \sqrt{x_1 \bar{x}_1 Q^2 \boldsymbol{z}_{13}^2 } \right) + \left( i\left( p_{M} - q \right)_{\perp \nu} - 2 z_{13 \perp \nu} \frac{i (q-p_M)_{\perp} \cdot z_{13 \perp} }{ \boldsymbol{z}_{13}^2 } \right) \right. \right. \nonumber \\ \left. \left. \times   \left( \frac{x_3}{(x_1+x_3) x_1 x_2} K_2 \left( \sqrt{ \frac{x_1 x_3 Q^2 \boldsymbol{z}_{13}^2}{x_1+x_3} } \right) - \frac{x_2 + x_3}{ x_1 x_2} K_2 \left( \sqrt{x_1 \bar{x}_1 Q^2 \boldsymbol{z}_{13}^2 } \right) \right)  \right] \right \}  \nonumber \\ \equiv -\frac{2 \pi}{x_2} \bigg \{ z_{13 \perp \nu} F_2 \left( x_1, x_3, \boldsymbol{z}_{13}^2 \right) +z_{13 \perp \nu} \; i (q-p_M)_{\perp} \cdot z_{13 \perp} F_2^{ \rm n. F.} \left( x_1, x_3, \boldsymbol{z}_{13}^2 \right) \nonumber \\ - ( \boldsymbol{z}_{13}^2 \; i (q-p_M)_{\perp \nu} + 2 z_{13 \perp \nu} \; i (q-p_M)_{\perp} \cdot z_{13 \perp} ) \tilde{F}_2^{ \rm n. F.} \left( x_1, x_3, \boldsymbol{z}_{13}^2 \right) \bigg \} \; ,
    \label{Eq:I_b2_Fin}
\end{gather}
\begin{gather}
    \frac{I_3^{(b)} ( \{ x \}, z_{13} )}{e^{ i \frac{x_2}{x_1 + x_3} (x_1 \boldsymbol{z}_1 + x_3 \boldsymbol{z}_3 ) (\boldsymbol{q} - \boldsymbol{p}_M)}} \bigg |_{\rm F.} = 
     - \frac{2 \pi}{x_1 x_2 Q^2 \boldsymbol{z}_{13}^2} \left(g_{\perp \alpha \nu}+\frac{2 z_{13 \perp \nu} z_{13 \perp \alpha}}{\boldsymbol{z}_{13}^2}\right)  \nonumber \\ \times \left[\sqrt{x_1 \bar{x}_1 Q^2 \boldsymbol{z}_{13}^2} K_1\left(\sqrt{x_1 \bar{x}_1 Q^2 \boldsymbol{z}_{13}^2}\right)-\sqrt{\frac{x_1 x_3}{x_1+x_3} Q^2 \boldsymbol{z}_{13}^2} K_1\left(\sqrt{\frac{x_1 x_3}{x_1+x_3} Q^2 \boldsymbol{z}_{13}^2}\right)\right]  \nonumber \\
    -  \frac{2 \pi}{x_1 x_2} \frac{\boldsymbol{z}_{13 \perp \nu} z_{13 \perp \alpha}}{\boldsymbol{z}_{13}^2} \left[x_1 \bar{x}_1 K_0\left(\sqrt{x_1 \bar{x}_1 Q^2 \boldsymbol{z}_{13}^2} \right) - \frac{x_1 x_3}{x_1+x_3} K_0\left(\sqrt{\frac{x_1 x_3}{x_1+x_3} Q^2 \boldsymbol{z}_{13}^2}\right)\right] \nonumber \\
   = - \frac{2 \pi}{x_2} \left[ \left(g_{\perp \alpha \nu}+\frac{2 z_{13 \perp \nu} z_{13 \perp \alpha}}{\boldsymbol{z}_{13}^2}\right) F_3 \left( x_1, x_3 , \boldsymbol{z}_{13}^2  \right) + \frac{\boldsymbol{z}_{13 \perp \nu} z_{13 \perp \alpha}}{\boldsymbol{z}_{13}^2} \tilde{F}_3 \left( x_1, x_3 , \boldsymbol{z}_{13}^2  \right) \right] \; ,
   \label{Eq:I_b3_Fin}
\end{gather}
\begin{gather}
    \frac{ I_4^{(b)} ( \{ x \}, z_{13} ) }{ e^{ i \frac{x_2}{x_1 + x_3} (x_1 \boldsymbol{z}_1 + x_3 \boldsymbol{z}_3 ) (\boldsymbol{q} - \boldsymbol{p}_M)} }  =  \frac{2 \pi}{x_2 (x_1 + x_3) Q^2} \sqrt{\frac{x_1 x_3}{x_1+x_3} \boldsymbol{z}_{13}^2 Q^2} \nonumber \\ \times \frac{K_1 \left(  \sqrt{ \frac{ x_1 x_3 }{ x_1 + x_3} Q^2 \boldsymbol{z}_{13}^2 } \sqrt{ 1 + \frac{x_2}{x_1+x_3} \frac{(\boldsymbol{q}-\boldsymbol{p}_M)^2}{Q^2} } \right)}{  \sqrt{ 1 + \frac{x_2}{x_1+x_3} \frac{(\boldsymbol{q}-\boldsymbol{p}_M)^2}{Q^2} }} 
    \equiv \frac{2 \pi}{x_2} \boldsymbol{z}_{13}^2 F_4 \left( x_1, x_3, \boldsymbol{z}_{13}^2 \right)  \; .
    \label{Eq:I_b4_Fin}
\end{gather}
Armed with these integrals and performing the change of variables\footnote{Please note that after this change of variable it is evident that the transformation $( x_1 \leftrightarrow x_2, \boldsymbol{z}_{13} \rightarrow - \boldsymbol{z}_{13})$ is equivalent to the $(1 \leftrightarrow 2)$ one.}
\begin{equation}
     \hspace{0.5 cm} \boldsymbol{r} = \boldsymbol{z}_{1} - \boldsymbol{z}_{3} \; , \hspace{1 cm} \boldsymbol{b} = \frac{x_1 \boldsymbol{z}_1 + x_3 \boldsymbol{z}_3}{x_1 + x_3} \; ,
     \label{Eq:Dipole_Impact_13}
\end{equation}
we get
\begin{gather}
  \mathcal{A}_3^{(b) T, L} =  (2 \pi)^4 W^2 \delta^{(4)} (q + p_t - p_M -p_{t'} ) \left( \prod_{i=1}^3 \int_0^1 d x_i \right) \delta \left(1- \sum_i^3 x_i \right) \int d^2 \boldsymbol{r} \; \mathcal{U} (\boldsymbol{r}) \nonumber \\ \times \left[ - \frac{e_q m_M}{ \pi} Q^2 (\boldsymbol{r} \cdot \boldsymbol{\Delta}) \right]  \left[  (\boldsymbol{\varepsilon}_M^{*} \cdot \boldsymbol{r}) \frac{\varepsilon_{\gamma}^{+}}{q^+} \right] \Bigg \{  c_f \phi_{3 L} (x_1, x_2)  \nonumber \\ \times   \left( \frac{x_1  F_2^{ \rm n. F.} (x_1, x_3, \boldsymbol{r}^2) }{x_2} - \frac{x_1^2 F_2 (x_1, x_3, \boldsymbol{r}) }{x_2 (x_1+x_3)} + \frac{x_2 F_4 \left( x_2, x_3 , \boldsymbol{r}^2 \right)}{x_2+x_3} \right)  + (x_1 \leftrightarrow x_2) \Bigg \} 
\end{gather}
and
\begin{gather}
  \mathcal{A}_3^{(b) T, T} =  (2 \pi)^4 W^2 \delta^{(4)} (q + p_t - p_M -p_{t'} ) \left( \prod_{i=1}^3 \int_0^1 d x_i \right) \delta \left( 1 - \sum_i^3 x_i \right) \int d^2 \boldsymbol{r} \; \mathcal{U} (\boldsymbol{r}) \nonumber \\ \times \left( - \frac{e_q m_M}{8 \pi} \right) c_f \Bigg \{ - \frac{8}{x_2 x_3} T_{\rm f.} (\boldsymbol{r}) \bigg(2 F_3 (x_1, x_3, \boldsymbol{r} ) + \tilde{F}_3 (x_1, x_3, \boldsymbol{r} ) \bigg) \phi_{3, (+) } (x_1,x_2) - 4 \; T_{\rm n. f.} (\boldsymbol{r}) \nonumber \\  \times \left[ \frac{x_1 x_2}{(1 - x_1)(1-x_2)}  F_1 (x_1, x_3, \boldsymbol{r} ) \left( \frac{\phi_{3, (+) } (x_1,x_2)}{x_2} - \frac{2}{x_2} f_{3M}^A A(x_1, x_2) \right) \right. \nonumber \\  \left.  + \frac{\tilde{F}_3 (x_1, x_3, \boldsymbol{r} )}{(1-x_1) x_2 x_3} \bigg( -x_1 \phi_{3, (+) } (x_1,x_2) + (1-2 x_1) \phi_{3, (-) } (x_1,x_2) \bigg) \right] + (x_1 \leftrightarrow x_2) \Bigg \} \; .
\end{gather}
In the present form, it is not immediate that the result can be expressed in terms of only the independent combinations of DAs $\phi_{3, (\pm)}$. However, by manipulating the proportional term to $A(x_1,x_2)$, as follows:
\begin{gather}
    \left( \prod_{i=1}^3 \int_0^1 d x_i \right) \delta \left( 1 - \sum_i^3 x_i \right) ... \left( \frac{x_1 x_2}{(1-x_1)(1-x_2)} \right) \left[ \frac{2 F_1 (x_1, x_3, \boldsymbol{r}^2) }{x_2} f_{3M}^A A(x_1, x_2) \right] \nonumber \\  = \left( \prod_{i=1}^3 \int_0^1 d x_i \right) \delta \left( 1 - \sum_i^3 x_i \right) ... \left( \frac{x_1 x_2}{(1-x_1)(1-x_2)} \right) \nonumber \\ \times \left[ - \frac{x_2 F_1 (x_2, x_3, \boldsymbol{r}^2) + x_1 F_1 (x_1, x_3, \boldsymbol{r}^2) }{x_1 x_2} \right] \phi_{3, (-)} (x_1,x_2)  \; ,
\end{gather}
we obtain 
\begin{gather}
  \mathcal{A}_3^{(b) T, T} =  (2 \pi)^4 W^2 \delta^{(4)} (q + p_t - p_M -p_{t'} ) \left( \prod_{i=1}^3 \int_0^1 d x_i \right) \delta \left( 1 - \sum_i^3 x_i \right) \int d^2 \boldsymbol{r} \; \mathcal{U} (\boldsymbol{r}) \nonumber \\ \times \left( - \frac{e_q m_M}{8 \pi} \right) c_f \Bigg \{ - \frac{8}{x_2 x_3} T_{\rm f.} (\boldsymbol{r}) \bigg(2 F_3 (x_1, x_3, \boldsymbol{r} ) + \tilde{F}_3 (x_1, x_3, \boldsymbol{r} ) \bigg) \phi_{3, (+) } (x_1,x_2) - 4 \; T_{\rm n. f.} (\boldsymbol{r}) \nonumber \\  \times \left[ \frac{x_1 x_2}{(1 - x_1)(1-x_2)} \left( \frac{\phi_{3, (+) } (x_1,x_2)}{x_2} F_1 (x_1, x_3, \boldsymbol{r} ) \right. \right. \nonumber \\ \left. \left. + \frac{x_2 F_1 (x_2, x_3, \boldsymbol{r}^2) + x_1 F_1 (x_1, x_3, \boldsymbol{r}^2) }{x_1 x_2}  \phi_{3, (-)} (x_1,x_2) \right) \right. \nonumber  \\ \left.  - \frac{\tilde{F}_3 (x_1, x_3, \boldsymbol{r} )}{(1-x_1) x_2 x_3} \bigg( x_1 \phi_{3, (+) } (x_1,x_2) - (1-2 x_1) \phi_{3, (-) } (x_1,x_2) \bigg) \right] + (x_1 \leftrightarrow x_2) \Bigg \} \; .
\end{gather}

\subsection{Full result for $\mathcal{A}_3^{T, L}$ and $\mathcal{A}_3^{T, T}$ }
Combining the results of Appendices \ref{sec:Uz1} and \ref{sec:Uz2z3}, for the longitudinal-photon-to-transverse-meson amplitude, we obtain
\begin{gather}
  \mathcal{A}_3^{T, L} =  (2 \pi)^4 W^2 \delta^{(4)} (q + p_t - p_M -p_{t'} ) \left( \prod_{i=1}^3 \int_0^1 d x_i \right) \delta \left(1- \sum_i^3 x_i \right) \int d^2 \boldsymbol{r} \; \mathcal{U} (\boldsymbol{r}) (\boldsymbol{r} \cdot \boldsymbol{\Delta})  \nonumber \\ \times  \tau_L  \left[  (\boldsymbol{\varepsilon}_M^{*} \cdot \boldsymbol{r}) \frac{\varepsilon_{\gamma}^{+}}{q^+} \right]  \Bigg \{ \phi_{3 L} (x_1, x_2) \left[ (1-c_f)   \bigg( \frac{x_1^2}{x_3 (x_1+x_2)} F_2 (x_1, x_2, \boldsymbol{r}^2)  -  \frac{x_1 F_2^{ \rm n. F.} (x_1, x_2, \boldsymbol{r}^2)}{x_3}    \bigg) \right. \nonumber \\ \left. + c_f     \left( \frac{x_1  F_2^{ \rm n. F.} (x_1, x_3, \boldsymbol{r}^2) }{x_2} - \frac{x_1^2 F_2 (x_1, x_3, \boldsymbol{r}^2) }{x_2 (x_1+x_3)} + \frac{x_2 F_4 \left( x_2, x_3 , \boldsymbol{r}^2 \right)}{x_2+x_3} \right) \right]  + (x_1 \leftrightarrow x_2) \Bigg \} \; ,
  \label{Eq:A-LT-AlmostFin}
  \end{gather}
where $\tau_L = - e_q m_M Q^2 / \pi$.
Similarly, the full result for the transverse-photon-to-transverse-meson amplitude reads
\begin{gather}
  \mathcal{A}_3^{  T, T} =  (2 \pi)^4 W^2 \delta^{(4)} (q + p_t - p_M -p_{t'} ) \left( \prod_{i=1}^3 \int_0^1 d x_i \right) \delta \left( 1 - \sum_i^3 x_i \right) \int d^2 \boldsymbol{r} \; \mathcal{U} (\boldsymbol{r}) \; \nonumber \\ \times  \tau_T \Bigg \{  T_{\rm f.} (\boldsymbol{r}) 2 \bigg( c_f \frac{2 F_3 (x_1, x_3, \boldsymbol{r} ) + \tilde{F}_3 (x_1, x_3, \boldsymbol{r} )}{x_2 x_3} - (1-c_f) \frac{2 F_3 (x_1, x_2, \boldsymbol{r} ) + \tilde{F}_3 (x_1, x_2, \boldsymbol{r} )}{x_3^2} \bigg) \nonumber \\  \times \phi_{3, (+) } (x_1,x_2)  + \; T_{\rm n. f.} (\boldsymbol{r})  \left[ \frac{x_1 x_2}{(1-x_2)} \left( c_f \left( \frac{\phi_{3, (+) } (x_1,x_2)}{ (1 - x_1) x_2 } F_1 (x_1, x_3, \boldsymbol{r} ) \right. \right. \right. \nonumber  \\ \left. \left. \left. + \frac{x_2 F_1 (x_2, x_3, \boldsymbol{r}^2) + x_1 F_1 (x_1, x_3, \boldsymbol{r}^2) }{ (1 - x_1) x_1 x_2}  \phi_{3, (-)} (x_1,x_2) \right) + (1-c_f) \frac{F_1 (x_1, x_2, \boldsymbol{r} )}{x_3} \phi_{3, (-)} (x_1,x_2)\right) \right. \nonumber \\ \left.  + \left( \frac{(1-c_f)\tilde{F}_3 (x_1, x_2, \boldsymbol{r} )}{ (1-x_1) x_3^2 } - \frac{c_f \tilde{F}_3 (x_1, x_3, \boldsymbol{r} )}{(1-x_1) x_3 x_2}  \right) \bigg( x_1 \phi_{3, (+) } (x_1,x_2) - (1-2 x_1) \phi_{3, (-) } (x_1,x_2) \bigg) \right] \nonumber \\ + (x_1 \leftrightarrow x_2) \Bigg \} \; ,
  \label{Eq:A-TT-AlmostFin}
\end{gather}
where $\tau_T = e_q m_M / 2 \pi$.

\subsection{Summary of $F_i (x_i,x_j,\boldsymbol{r}^2)$ and $\tilde{F}_i (x_i,x_j,\boldsymbol{r}^2)$'s functions}
\label{App:F_i_functions}
For the convenience of the reader, we report here all the $F_i (x_i,x_j,\boldsymbol{r}^2)$ and $\tilde{F}_i (x_i,x_j,\boldsymbol{r}^2)$ functions appearing in the final results of the previous section:
\begin{gather}
    F_1 \left( x_i, x_j, \boldsymbol{r}^2 \right) = \frac{1}{(x_i + x_j)} K_0 \left(  \sqrt{ \frac{ x_i x_j }{ x_i + x_j } Q^2 \boldsymbol{r}^2 } \right) 
     \; ,
\end{gather}
\begin{gather}
    F_2 \left( x_i, x_j, \boldsymbol{r}^2 \right) = \sqrt{ \frac{ \bar{x}_i }{x_iQ^2 \boldsymbol{r}^2  } } K_1 \left( \sqrt{x_i \bar{x}_i Q^2 \boldsymbol{r}^2 } \right) - \sqrt{ \frac{ x_j }{x_i (x_i + x_j) Q^2 \boldsymbol{r}^2 }  } \; K_1 \left(  \sqrt{ \frac{x_i x_j }{x_i + x_j} Q^2 \boldsymbol{r}^2 }  \right)
     \; ,
\end{gather}
\begin{gather}
    F_2^{ \rm n.F.} \left( x_i, x_j, \boldsymbol{r}^2 \right) = \frac{(1-x_i-x_j)}{(x_i+x_j)}   \sqrt{ \frac{x_i \bar{x}_i}{ Q^2 \boldsymbol{r}^2} }   K_1 \left( \sqrt{x_i \bar{x}_i Q^2 \boldsymbol{r}^2 } \right) \; ,
\end{gather}
\begin{gather}
    \tilde{F}_2^{ \rm n.F.} \left( x_i, x_j, \boldsymbol{r}^2 \right) = \frac{1}{Q^2 \boldsymbol{r}^2  x_i}   \left( \frac{x_j}{(x_i + x_j) } K_2 \left( \sqrt{ \frac{x_i x_j Q^2 \boldsymbol{r}^2}{x_i+x_j} } \right) -  (1 - x_i) K_2 \left( \sqrt{x_i \bar{x}_i Q^2 \boldsymbol{r}^2 } \right) \right) \; ,
\end{gather}
\begin{gather}
    F_3 \left( x_i, x_j , \boldsymbol{r}^2 \right) = \sqrt{ \frac{\bar{x}_i}{x_i Q^2 \boldsymbol{r}^2} } K_1\left(\sqrt{ x_i \bar{x}_i Q^2 \boldsymbol{r}^2}\right) - \sqrt{\frac{ x_j}{x_i (x_i + x_j) Q^2 \boldsymbol{r}^2} } K_1\left(\sqrt{\frac{x_i x_j }{x_i + x_j } Q^2 \boldsymbol{r}^2}\right)    ,
\end{gather}
\begin{gather}
    \tilde{F}_3 \left( x_i, x_j , \boldsymbol{r}^2  \right) = \bar{x}_i K_0\left(\sqrt{x_i \bar{x}_i Q^2 \boldsymbol{r}^2}\right) - \frac{x_j}{x_i + x_j } K_0\left(\sqrt{\frac{x_i x_j}{ x_i + x_j } Q^2 \boldsymbol{r}^2}\right)  
\end{gather}
and
\begin{gather}
     F_4 \left( x_i, x_j, \boldsymbol{r}^2 \right) =  \frac{1}{ (x_i + x_j ) Q^2 \boldsymbol{r}^2} \sqrt{\frac{x_i x_j}{(x_i + x_j)} Q^2 \boldsymbol{r}^2 } \nonumber \\ \times \frac{K_1 \left(  \sqrt{ \frac{ x_i x_j }{ (x_i + x_j) } Q^2 \boldsymbol{r}^2 } \sqrt{ 1 + \frac{(1-x_i-x_j)}{(x_i+x_j)} \frac{\Delta^2}{Q^2} } \right)}{  \sqrt{ 1 + \frac{(1-x_i-x_j)}{(x_i+x_j)} \frac{\Delta^2}{Q^2} }} = \frac{1}{ (x_i + x_j ) Q^2 \boldsymbol{r}^2} \sqrt{\frac{x_i x_j}{(x_i + x_j)} Q^2 \boldsymbol{r}^2 } \nonumber \\ \times K_1 \left(  \sqrt{ \frac{ x_i x_j }{ (x_i + x_j) } Q^2 \boldsymbol{r}^2 }  \right) + \mathcal{O} (\Delta^2) \; .
\end{gather}

\newpage
\bibliographystyle{apsrev}
\bibliography{references}

\end{document}